\documentstyle[11pt,aaspp4] {article}
 
\def\etal{et al.}
\def\figa{({\it a})}
\def\figb{({\it b})}
\def\cf{c.f.}
\def\ie{i.e.}
\def\deg{^\circ}
\def\kms{\mbox{km/s}}
\def\hMpc{h^{-1}\mbox{Mpc}}
\def\nusig{\nu_\sigma}

\begin{document}

\title{Topology from the Simulated Sloan Digital Sky Survey}

\author{
Wesley N. Colley\altaffilmark{1},
J. Richard Gott, III\altaffilmark{2},
David H. Weinberg\altaffilmark{3},
Changbom Park\altaffilmark{4} and
Andreas A. Berlind\altaffilmark{3}
}
\altaffiltext{1}{Harvard-Smithsonian Center for Astrophysics, 60 Garden Street,
Cambridge MA 02138, wcolley@cfa.harvard.edu}

\altaffiltext{2}{Princeton University Department of Astrophysical Sciences,
Princeton, NJ 08544, jrg@astro.princeton.edu} 

\altaffiltext{3}{Department of Astronomy, The Ohio State University, 174 W 18th
Ave., Columbus, OH 43210-1106, dhw,aberlind@astronomy.ohio-state.edu}

\altaffiltext{4}{Department of Astronomy, Seoul National University, Seoul 151,
Korea, cbp@astro.snu.ac.kr}

\begin{abstract}

We measure the topology (genus curve) of the galaxy distribution in a very
large cosmological simulation designed to resemble closely results from the
upcoming Sloan Digital Sky Survey (SDSS).  The mock survey is based on a large
N-body simulation that uses 54,872,000 particles in a periodic cube $600\hMpc$
on a side.  The adopted inflationary cold dark matter (CDM) model has
parameters $\Omega_{\rm CDM}=0.4$, $\Omega_\Lambda=0.6$, $h=0.6$, $b=1.3$.  We
``observe'' this simulation to produce a simulated redshift catalog of $\sim
10^6$ galaxies over $\pi$ steradians, mimicking the ancitipated spectroscopic
selection procedures of the SDSS in some detail.  Sky maps, redshift slices,
and 3-D contour maps of the mock survey reveal a rich and complex structure,
including networks of voids and superclusters that resemble the patterns seen
in the CfA redshift survey and the Las Campanas Redshift Survey (LCRS).  The
3-D genus curve can be measured from the simulated catalog with superb
precision; this curve not only has the general shape predicted for Gaussian,
random phase initial conditions, but the error bars are small enough to
demonstrate with high significance the subtle departures from this shape caused
by non-linear gravitational evolution on a $10\hMpc$ smoothing scale (where
$\sigma_{\rm gal}=0.4$).  These distortions have the form predicted by
Matsubara's (1994) perturbative analysis, but they are smaller in amplitude.
We also measure the 3-D genus curve of the radial peculiar velocity field
measured by applying distance-indicator relations (with realistic errors) to
the mock catalog.  The genus curve is consistent with the Gaussian random phase
prediction, though it is of relatively low precision because of the large
smoothing length required to overcome noise in the measured velocity field.
Finally, we measure the 2-D topology in redshift slices, similar to those which
will become available early in the course of the SDSS and to those already
observed in the LCRS.  The genus curves of these slices are consistent with the
observed genus curves of the LCRS, suggesting that our inflationary CDM model
with $\Omega_{\rm CDM}\sim 0.4$ is a good choice.  Our mock redshift catalog is
publicly available for the use of other researchers.

\end{abstract}

\section{Introduction}

The nature of large-scale structure in the universe is one of the pre-eminent
questions of modern astronomy.  Gravitational growth of structure from quantum
fluctuations during inflation has become the leading candidate for production
of this structure.  The topology of large-scale structure as quantified by the
``genus curve'' (Gott, Melott, \& Dickinson 1986, hereafter GMD; Hamilton,
Gott, \& Weinberg 1986; Gott, Weinberg, \& Melott 1987, hereafter GWM) tests
one of the most robust predictions of inflation, namely that the quantum
fluctuations produce a Gaussian random phase field of density perturbations,
the seeds of the structures seen today.  To date, all topological studies of
galaxy redshift surveys (in 3-D and 2-D) and cosmic microwave background (CMB)
anisotropies (in 2-D) have been consistent with Gaussian random phase initial
conditions (Gott \etal\ 1989; Moore \etal\ 1992; Park, Gott, \& da Costa 1992;
Gott, Rhoads, \& Postman 1994; Vogeley et al.\ 1994; Colley, Gott, \& Park
1996; Kogut et al.\ 1996; Colley 1997; Protogeros \& Weinberg 1997; Canavezes
et al.\ 1998; Springel et al.\ 1998).

The genus test becomes only more powerful as cosmological surveys become
larger.  While several prodigious observational efforts have been made to
survey the large-scale structure traced by optical galaxies (Geller \& Huchra
1989; Shectman \etal\ 1996) and by IRAS galaxies (Canavezes et al.\ 1998 and
references therein), the 2-degree Field (2dF) redshift survey (see Colless
1998) and the Sloan Digital Sky Survey (SDSS)\footnote{Detailed information
about the SDSS can be found in the collaboration's NASA proposal, available at
{\tt http://www.astro.princeton.edu/BBOOK/}, which we refer to henceforth as
Knapp et al.\ (1997).  Review articles describing the survey in more general
terms include Gunn \& Knapp (1993), Gunn \& Weinberg (1995), and Margon
(1999).}  will dwarf all currently existing redshift surveys in data volume.
The SDSS will obtain approximately one million galaxy redshifts (compared to
$\sim 250,000$ anticipated for 2dF), and the resulting high-precision
measurement of the three-dimensional topology will therefore provide a far more
powerful test of the random phase hypothesis than is possible today.  In order
to prepare for topological analysis and other statistical studies of this
enormous sample, we have created a simulation of the SDSS using a very large
N-body simulation ($N = 54,827,000$ particles).  In this paper we show that the
three-dimensional genus curve can be measured with unprecedented precision from
the SDSS, as anticipated.  We also show that it should be possible to measure
the topology of the smoothed radial velocity field inferred from the SDSS
images and spectra via distance-indicator relations.

Early on in the SDSS, two-dimensional redshift slices, comparable to the
deepest existing wide-angle redshift surveys, such as the Las Campanas Redshift
Survey (LCRS, Shectman \etal\ 1996), will allow studies of the two-dimensional
genus curve.  Colley (1997) computed the two-dimensional topology of
large-scale structure observed in the LCRS.  We have therefore generated slices
from our Simulated SDSS (SSDSS), intended to match closely those of the LCRS.
We use particular statistical care to compare the topology of our simulation
slices to that of the observed slices from the LCRS.  Redshift slices (of
constant angle in the sky) are complementary to maps of the CMB, which mainly
probe a shell at constant radius ($z \sim 1000$).  The topology of CMB
fluctuations as observed by COBE (Smoot \etal\ 1994) has been computed by
Colley \etal\ (1996) and Kogut et al. (1996).  A similar 2-D topological
approach has been applied to projected galaxy and cluster catalogs by Gott,
Mao, Park \& Lahav (1992), Plionis, Valdarnini, \& Coles (1992), Coles et al.\
(1993), and Davies \& Coles (1993).

\section{The Cosmological Simulation}

The SDSS will measure the redshifts of approximately $10^6$ galaxies in $\pi$
steradians of the northern sky.  This observational sample should provide an
unprecedented opportunity to measure quantitatively the 3-D and 2-D topology of
large-scale structure in the universe.  

Since, at its best, theory should predict observations {\it before} they are
taken, we felt it important to simulate {\it ahead} of time the results to be
obtained by the SDSS.  One purpose of this is cosmological model testing, of
course, but no less importantly we hope to show in a general way what features
may be expected from a gravitational instability model.  Perhaps most important
of all, we hope to illustrate just how powerful the SDSS will be, i.e., just
how accurately it will allow us to measure statistical quantities of interest,
such as the genus curve.  To this end, we have run a very large N-body
simulation intended to mimick results from the SDSS.

In 1992 the results from COBE (Smoot \etal\ 1992; Wright \etal\ 1992) provided
great support for the gravitational instability picture by finally showing the
long sought fluctuations in the microwave background predicted by this theory.
The fluctuations found by COBE were consistent with a Harrison (1970)-Peebles
\& Yu (1970)-Zeldovich (1972) spectrum of fluctuations with a primordial index
$n = 1$, Gaussian random phases (e.g. Colley et al.\ 1996; Kogut et al.\ 1996),
and a standard, unbiased ($b \approx 1$), $\Omega = 1$, $h = 0.5$ (where
$h\equiv H_0/[100~\mbox{km/s/Mpc}]$), inflationary cold dark matter (CDM)
universe.  It is noteworthy that the fluctuations seen by COBE are on scales
larger than the horizon size at recombination, just the sort of fluctuations
one would expect from random quantum fluctuations in a standard inflationary
model (Guth 1981).

While CDM inflationary models allow low amplitude microwave background
fluctuations like those observed by COBE, they simultaneously produce structure
which matches that observed at redshifts $z \la 5$, because the smoothly
distributed baryons can fall into pre-existing CDM potential wells {\it after}
recombination and because the hierarchical form of the inflationary CDM power
spectrum (Bardeen, Steinhardt, \& Turner 1983) allows early formation of
quasars and galaxies.  Furthermore, adiabatic Gaussian fluctuations with this
initial $n = 1$ power spectrum have proven amazingly successful at explaining
the qualitative features of observed galaxy clustering, including the
characteristic pattern of great walls and giant voids seen in the Geller \&
Huchra (1989) slices (White et al.\ 1987; Park 1990; Weinberg \& Gunn 1990a)
and lattice-like sequences of great walls in deep pencil beam surveys
(Broadhurst \etal\ 1990; Park \& Gott 1991a).  Other successes of CDM include
this very wide variety of remarkable properties in common with observation:
deep sky maps show clusters, but are bland compared to $6^\circ$ slices which
show large voids; $20^\circ$ wide slices show great walls; velocity fields show
great attractors; the 3-D topology is spongelike; deep pencil beams show
surprisingly regular lattices of great walls out to $z = 0.5$; deep slices to
$z = 0.2$ show a more uniform appearance than shallow slices to $z = 0.05$; the
cluster-cluster covariance function is a higher amplitude version of the
galaxy-galaxy covariance function (Bahcall \& Soneira 1983; Klypin \& Kopylov
1983); and the three-point correlation functions of clusters and galaxies show
the same triangular form (Gott, Gao, \& Park 1991).  Even if one were given
complete freedom to construct an arbitrary, multi-parameter, geometrical model
of large-scale structure, it is hard to see how one could come up with
something that would reproduce all of these qualitative and quantitative
successes of the physically motivated, dynamical model of gravitational
structure formation from Gaussian initial conditions with an inflationary CDM
power spectrum.

Within the inflationary CDM paradigm, there are many reasons for selecting a
model with $\Omega \approx 0.4$.  In models with scale-invariant inflationary
fluctuations ($n=1$), CDM-dominated matter content ($\Omega_{\rm HDM} \ll
\Omega_{\rm CDM}$; $\Omega_B \ll \Omega_{\rm CDM}$), and a standard
relativistic background (CMB photons plus the usual light neutrino species), a
low value of $\Omega h$ is required to explain the observed shape of the galaxy
power spectrum (see Peacock \& Dodds 1994 and numerous references therein) and
the observed amplitude of the 3-D galaxy genus curve at large smoothing scales
(see Moore et al.\ 1992; Vogeley et al.\ 1994; Canavezes et al.\ 1998).  With
the same restrictions, a value of $\Omega \sim 0.25-0.5$ is required to match
simultaneously the COBE fluctuation amplitude and the mass function of galaxy
clusters for $t_0 \approx 13$ Gyr (Cole et al.\ 1997 and references therein).
Direct evidence for $\Omega \approx 0.4$ includes the combination of the
cluster baryon fraction with big bang nucleosynthesis constraints (Evrard 1997)
and the combination of the $z=2.5$ mass power spectrum inferred from the
Lyman-alpha forest with the $z=0$ cluster mass function (Weinberg et al.\
1998).  This value of $\Omega$ is consistent with some analyses of galaxy
cluster evolution (e.g., Bahcall, Fan, \& Cen 1997; Eke, Cole, \& Frenk 1998)
but not with others (e.g., Blanchard \& Bartlett 1998; Reichart et al.\ 1998;
Viana \& Liddle 1998).  With a modest bias of the galaxy population in clusters
it is consistent with observed cluster mass-to-light ratios (Carlberg et al.\
1996, 1997).

Once one adopts low $\Omega$, the inclusion of a cosmological constant
$\Omega_\Lambda=1-\Omega$ is theoretically attractive (Ratra \& Peebles 1998;
Peebles \& Ratra 1998) because it maintains the flat spatial geometry that
seems the most natural prediction of inflation, requiring no fine-tuning of the
number of inflationary $e$-folds (though fine-tuning is still required to
explain the present-day value of $\Omega_\Lambda$).  Such a model could arise
naturally within Linde's (1990) chaotic inflation scenario.  Flat geometry also
appears to be favored by some preliminary measurements of the angular location
of the first Doppler peak in the CMB multipole spectrum (see, e.g., discussions
by Lineweaver [1998] and Tegmark [1998]).  A cosmological constant makes it
easy to reconcile values of $h \sim 0.6-0.8$ with the inferred ages of the
oldest globular clusters.  A value of $\Omega_\Lambda=0.6$ is low enough that
gravitational lensing cross-sections are not an embarrassment (Fukugita \etal\
1992), and indeed Chiba \& Yoshii (1999) argue that gravitational lensing
statistics favor a value of $\Omega_\Lambda$ in this range.  The recent
inference of an accelerating cosmic expansion from observations of Type Ia
supernovae (Riess et al.\ 1998; Perlmutter et al.\ 1999) provides additional
and more direct evidence for a cosmological constant or some similar
negative-pressure component of the universe.  These and other arguments in
favor of a low-density, flat universe have been summarized by a number of
authors, including a recent analysis by Roos \& Harun-or-Rashid (1999), who
show that the combination of a variety of independent constraints favors
$\Omega_\Lambda=0.70\pm 0.14$ and $\Omega_{\rm M}+\Omega_\Lambda=0.99 \pm 0.16$
($1\sigma$ error bars) even {\it without} the supernova data.

An alternative to the $\Lambda\mbox{CDM}$ model is an open model with
$\Omega_{\rm CDM}<1$ and $k=-1$.  An open universe can be produced in
single-bubble inflationary models (cf. Gott 1982; Gott \& Statler 1984; Gott
1986; Ratra \& Peebles 1994, 1995; Bucher \etal\ 1995a,b; Yamamoto \etal\ 1995;
Linde 1995; Linde \& Mezhlnmian 1995; Hawking \& Turok 1998).  An $\Omega =
0.4$, $\Omega_\Lambda = 0.6$, $b = 1.3$ model is quite similar, in terms of
galaxy clustering, to an $\Omega = 0.4$, $\Omega_\Lambda = 0$, $b = 1.3$ model,
so our simulation can serve reasonably well as a stand-in for either.

Ratra \etal\ (private communication) find the following 4-year-COBE normalized
values with 2-$\sigma$ limits, all assuming $\Omega _{B}h^2 = 0.0125$:
\begin{equation}
\label{eqn_cosmparams}
\begin{array}{ll}
\\
\Omega =1, \Omega_\Lambda=0, h=0.54, t_0= 12~\mbox{Gyr},\\ 
\Omega h=0.54, \sigma_8 = 1.1\mbox{--}1.5,
0.7 < b <0.9;\\ 
\\
\Omega =0.4, \Omega_\Lambda = 0.6, h=0.6, t_0= 14.5~\mbox{Gyr},\\
\Omega h=0.24, \sigma_8 =0.82\mbox{--}1.2,
0.8 < b <1.22;\\ 
\\ 
\Omega =0.4, \Omega_\Lambda=0, h=0.63, t_0 = 12~\mbox{Gyr}, \\
\Omega h=0.252, \sigma_8 =0.51\mbox{--}0.74,
1.35 < b <1.96.
\end{array}
\end{equation}
Thus our
simulation's bias parameter $b = 1.3$ (our simulation was done before the COBE
4-year normalization was available) is just above the $0.8 < b < 1.22$ COBE
4-year normalization, so it has only marginally high bias.  Its bias is
slightly below the COBE 4-year normalization for an $\Omega = 0.4$,
$\Omega_\Lambda = 0$ model ($1.35 < b < 1.96$).
This modest level of bias is consistent with the predictions of
hydrodynamic cosmological simulations (e.g., Cen \& Ostriker 1992;
Katz, Hernquist, \& Weinberg 1992) and semi-analytic models of
galaxy formation (e.g., Kauffmann, Nusser, \& Steinmetz 1997).

We thus adopt a CDM inflationary model with cosmological parameter values
$\Omega = 0.4$, $\Omega_\Lambda = 0.6$, $k = 0$, $h = 0.6$, $t_0 =
14.5~\mbox{Gyr}$, and r.m.s. fractional mass fluctuation $\sigma_8=1/b=0.77$ in
spheres of radius $8\hMpc$.  These values arguably provide the best fit to all
current observational constraints. (Ostriker \& Steinhardt [1995] independently
picked as their favored model one with similar parameters; Turner [1998] also
independently arrived at these as a promising set of parameters, roughly
consistent with recent Type Ia supernovae data from Reiss \etal\ [1998] and
Perlmutter \etal\ [1999], which favor $\Omega \sim 0.2\mbox{--}0.3$.)

Our N-body simulation uses $N = 380^3 = 54,872,000$ particles (the simulation
was run in 1993; some results from it were published in Vogeley \etal\ [1994],
prior to the appearance of this paper).  The simulation uses a staggered-grid,
particle-mesh code (Park 1990) with a $600^3$ density-potential mesh and a
simulation volume of $(600\hMpc)^3$.  Thus our gravitational force resolution
is $\sim 1\hMpc$.  A biased subset of 8,292,455 particles are chosen to
represent galaxies, chosen as peaks that lie above a threshold $\delta /
\delta_{rms} = 0.8$ when the CDM density field is smoothed over $0.71\hMpc$.
The peak particles are identified using the Bardeen et al.\ (1986)
peak-background split approximation, as discussed by Park (1991).  This
combination of smoothing scale and peak threshold yields a bias factor $b =
1.3$ between the r.m.s. fluctuations of galaxy particles and dark matter on
large scales.

\section{The Mock Redshift Catalog}

We have attempted to model the anticipated selection properties of the
SDSS redshift survey in some detail, in part because our simulated
redshift survey is being used for a number of internal tests of the
survey data analysis software and observing strategy.  The present
observational plan
is to select galaxies in the main redshift sample based on their
Petrosian (1976) magnitudes and half-light surface brightnesses 
in the $r^\prime$ band (see Fukugita et al.\ [1996] and Gunn et al.\ [1998]
for discussions of the SDSS photometric system).
Definitions of our Petrosian magnitude system and motivation for its
use are given by Gunn \& Weinberg (1995).  In the notation of that
paper we adopt parameters $f_1=1/8$ and $f_2=2$ for our mock redshift
catalog, which means that we define magnitudes within a circular
aperture of radius $2R_P$, where the Petrosian radius $R_P$ is the
radius at which the local surface brightness falls to $1/8$ of the
mean interior surface brightness.  As a result of more recent tests,
the SDSS is likely to adopt somewhat different values of $f_1$ and
$f_2$, but we do not expect these changes to substantially alter
the survey selection function, since the magnitude limit will be
adjusted to ensure the desired number of galaxy targets over the
$\pi$-steradian survey area.

In order to compute the quantities used for target selection, we first
assign two fundamental parameters to each of the 8,292,455 galaxy
particles: a $B_T$ luminosity and a Hubble type.
The $B_T$ luminosities are randomly drawn from a Schechter (1976)
luminosity function with parameters $M_*=-19.68+5\log h$, $\alpha=-1.07$,
truncated below 0.064$L_*$ (about 3 magnitudes below $M_*$) so that
the space density of galaxies above the cutoff matches the
space density ${\bar n}=0.038h^3{\rm Mpc}^{-3}$ of galaxies in
the N-body simulation.  Hubble types E--Sc are assigned randomly,
with probabilities modified according to local galaxy density
in accord with the Postman \& Geller (1984) morphology-density relation
(see Narayanan, Berlind, \& Weinberg [1998] for details).
Bulge-to-disk ratios are assigned as a function of Hubble type
based on Kent (1985), and bulge axis ratios are drawn from the
distribution inferred by Ryden (1992).  Spheroid half-light radii are
assigned using Maoz \& Rix's (1992) parametrization of the
fundamental plane (Djorgovski \& Davis 1985; Dressler et al.\ 1987)
for ellipticals, modified according to Fig.~5\figb~of Kent (1985)
for spiral bulges. Disk half-light radii are assigned using 
Freeman's (1970) value for the typical central surface brightness.
We add Gaussian scatter with r.m.s. of 0.15 in $\log_{10}D$ to 
the diameters $D$ in order to ensure a population of high and low
surface brightness galaxies; this scatter may be unrealistically
large, at least for the bulge components.  We compute K-corrected
magnitudes in the SDSS filters using the spectral-energy distributions
of Coleman, Wu, \& Weedman (1980).  Finally, we compute Petrosian
magnitudes and half-light surface brightnesses assuming a 
de Vaucouleurs (1948) profile for the bulge components and an
inclined exponential profile for the disk components.

We adopt a half-light
surface brightness threshold $\mu_{1/2}=22$ mag arcsec$^{-2}$
and a Petrosian magnitude limit $r^\prime=17.9$ (on the AB magnitude
system, Oke \& Gunn [1983]), yielding 872,377 galaxies in the mock redshift
catalog.  The SDSS currently plans to target $\sim 10^5$ luminous red
ellipticals in addition to the main galaxy sample, using photometric
redshifts to obtain a sparse, nearly volume-limited sample extending to 
$z \sim 0.4$.  We attempt to model only the $\sim 9\times 10^5$ 
main galaxy sample here because of the size of our simulation cube.

Once we have decided which galaxies are bright enough to be included in the
mock catalog (with origin at a corner of the periodic simulation cube), we need
to set the survey
sky coverage.  The north Galactic cap portion of the SDSS will target a roughly
elliptical region, $130^\circ$ by $110^\circ$.  One may construct this region
by choosing a polar coordinate system ($\theta ,\phi $) centered on the center
of the ellipse.  The two foci A and B of the ellipse are along the major axis,
each at a distance $\theta_F = 42.54\deg =
\cos^{-1}[\cos(130\deg/2)/\cos(110\deg/2)]$.  Let ($\theta_g,\phi_g$) be the
polar coordinates of a galaxy in the sky.  Its angular distances in the sky
from foci A and B respectively are:
\begin{equation}
\label{eqn_surveyarea1}
\begin{array}{ll}
\theta_{gA} = \cos^{-1} [\sin \theta_F\sin\theta_g\cos\phi_g
+ \cos\theta_g \cos\theta_F] \\

\theta_{gB} = \cos^{-1} [-\sin \theta_F \sin \theta_g\cos\phi_g
+ \cos\theta_g \cos \theta_F].
\end{array}
\end{equation}
The galaxy is included in the elliptical survey region if
\begin{equation}
\label{eqn_surveyarea2}
\theta_{gA} + \theta_{gB}\leq 130\deg
\end{equation}
Fig.~\ref{skyfig} shows a sky map of the $\sim 900,000$ galaxies in the
survey region.  
This figure bears a remarkable qualitative resemblance to 
the map of the real sky derived from the Shane-Wirtanen (1967) counts
(reproduced on page 41 of Peebles [1993]), which reach to similar depth
(without redshifts, of course).

Fig.~\ref{bigslice} shows how a $6\deg\times 130\deg$ slice along the major
axis of the survey looks when the observed galaxies are plotted in redshift
space.  Galaxy redshifts are converted into comoving distances using the
redshift-comoving distance relation of the $\Omega_\Lambda=0.6$, $\Omega=0.4$
cosmological model.  (A simple cubic approximation to this relation,
$r_c=3000z-940z^2+130z^3\;\hMpc$, has a maximum error of $0.46\hMpc$ out to
$z=0.4$.)  The observer is located at the vertex of the fan, which extends to a
redshift corresponding to a comoving distance of $500\hMpc$.  Because we have a
lower luminosity limit for galaxies (3 magnitudes below $M_*$ in $B$), this
magnitude-limited sample is incomplete at distances below $\sim 130\hMpc$; the
combination of this limit with the decreasing physical width of the slice
causes the apparent deficit of galaxies near the vertex of the fan.  In the
region from $130\hMpc$ to $400\hMpc$ where the density of galaxies is highest
and the clustering is most easily seen, we find a wealth of structure.  The
many small ``fingers of god'' pointing toward the observer show the locations
of clusters; each ``finger'' corresponds to a roughly spherical cluster, which
is stretched in redshift space by the peculiar velocities of its member
galaxies.  Many voids can be seen with typical sizes of $30\hMpc$ to $50\hMpc$,
quite similar to those observed in the de Lapparent, Geller \& Huchra (1986)
6$\deg$ slice, which goes out to $150\hMpc$.  This pattern of voids is quite
typical of CDM models (cf. Park 1990).  Many small and ``great'' walls (Geller
\& Huchra 1989) are visible, frequently up to $150\hMpc$ in length,
corresponding to the ``pancake'' structures that Zel'dovich (1970) argued would
be a natural consequence of gravitational clustering from Gaussian initial
conditions (see also Shandarin \& Zel'dovich 1989).  There is even a ``great
wall complex'' extending for $\sim 400\hMpc$ across the slice, at distances
$\sim 150-300\hMpc$ from the observer.  Such features are also seen in
$\Omega = 1$, $h = 0.5$, $b = 2$ CDM models (Park 1990).  The visual difference
between this model with $\Omega h = 0.24$ and a standard $\Omega h = 0.5$ CDM
model is in the low amplitude, rolling hills and valleys in the distribution.
The extra modulation of the basic network of voids and walls reflects the extra
power on large scales present in an $\Omega h = 0.24$ CDM model.  Nonetheless,
we can see that the structure in the slice is approaching a qualitatively
``fair sample'' of this simulated universe: this fan has a much more uniform
appearance than the de Lapparent et al.\ (1986) $6\deg$ slice, which only went
out to $150\hMpc$.  Any theory based on Friedmann cosmological models must
approach uniformity on large scales.  In the $\Omega h = 0.24$ CDM model, the
power spectrum $P(k)$ peaks at wavelength $\lambda \sim 260\hMpc$, and it drops
at larger scales, approaching $P(k) \propto k$ at very long wavelengths.

In addition to
large voids $30\mbox{--}50\hMpc$ across, there are some void complexes
approximately $100\hMpc$ across (like the one at 1 o'clock at $250\hMpc$
distance), where we have several low density voids next to each other
separated by only relatively weak walls.  Weinberg \& Gunn (1990b)
show time sequences of the formation of such void complexes in models 
with $k^{-1}$ power spectra.  
Galaxies drain off the walls between voids, and the walls
eventually become so tenuous that they are barely noticeable.  
As pointed out by Park \etal\ (1992),
such a case of a tenuous wall inside a large void is seen in the
famous de Lapparent et al.\ (1986) slice.
As a CDM model evolves in time,  voids grow by galaxies
flowing off of minor caustics onto major caustics.  This is also the mechanism
for the enhancement of great walls.

Overall this slice, which goes to a depth of $500\hMpc$, looks remarkably like
the sandwich of three $1.5\deg$ wide slices (with two $1.5\deg$ wide gaps) from
the LCRS (Shectman et al.\ 1996).  Since our simulation was run before
completion of the LCRS, we were delighted to see how the LCRS qualitatively
confirmed many of the features of our simulation slice.  It shows the
same wall and void structure, the same rolling density at large scales, the
same void and wall complexes, and the same overall approach to uniformity on
very large scales.  The LCRS shows a power spectrum consistent with CDM and
$0.2 < \Omega h < 0.3$ (Lin et al.\ 1996) and a 2-D topology consistent with
random phase initial conditions (Colley 1997), as assumed in our simulation.

\section{Measuring the 3-D Topology of Large-Scale Structure}

Figs.~\ref{SSDSS.3d50}\figa~and \ref{SSDSS.3d50}\figb~show contours of constant density in a subset of the three-dimensional
simulation out to $500\hMpc$.  We have created a volume-limited sample for 
these figures by throwing out those galaxies that would not be visible at
$500\hMpc$.  After smoothing with a Gaussian filter of radius
$R_s= 10\hMpc$, we have selected contour surfaces at
the median density contour [\ref{SSDSS.3d50}({\it a})], and at the 
93rd-percentile density contour [\ref{SSDSS.3d50}({\it b})] 
(such that the 7\% of the volume with highest density 
is contained within this contour).  The median
density contour surface is sponge-like, as expected from Gaussian random phase
initial conditions (GMD), while Fig.
\ref{SSDSS.3d50}\figb~shows isolated clusters at the 93rd-percentile 
density, as expected for this contour (GWM).
These figures illustrate the hundreds of structures that will be available to
constrain the 3-D genus in the complete SDSS.

In order to quantify the topology of the three-dimensional large-scale
structure in our mock redshift catalog, we follow the approach outlined
by GMD and GWM.  We start with the galaxy density
field ($\delta$-functions at the location of galaxies) and smooth it with a
smoothing function $W(r) = e ^{-r^2/2R_s^2}$, with $R_s$ larger than the
mean interparticle separation, 
to produce a smoothed density field.  Isodensity contour
surfaces can then be constructed, and the genus $G_{3D}$
of such a surface can be defined as
\begin{equation}
\label{eqn_genusdef}
G_{3D} = \mbox{No. of Holes} - \mbox{No. of isolated regions},
\end{equation}
where ``hole'' means hole like a donut has and ``isolated region'' refers to
a topologically separate, isolated structure.  For example, a surface
density contour surface which consisted of 10 spherical pieces each
surrounding an isolated spherical cluster would have a genus of $G_{3D} = -10$.
GMD prove that
\begin{equation}
\label{eqn_genuscurv}
G_{3D} = -{1\over{4\pi}}\int KdA,
\end{equation}
where $K$ is the Gaussian curvature ($= 1 / r_1 r_2$ where $r_1$ and $r_2$ are
the two principal radii of curvature) and the integral is performed over the
contour surface.  Contour surfaces can be labeled by the $\nu$ value, which is
related to the volume fraction of space $f$ that they enclose:
\begin{equation}
\label{eqn_gen_nudef}
f = {1\over{(2\pi)^{1/2}}}
\int_{\nu}^{\infty} e^{-t^2/2}dt.
\end{equation}
In the case of a Gaussian one-point probability distribution, the
value of $\nu$ in equation~(\ref{eqn_gen_nudef}) is equal to $\delta/\sigma$,
the density contrast in units of the standard deviation.  The definition of
contours in terms of fractional volume rather than density contrast greatly
reduces the influence of non-linear gravitational evolution and biased 
galaxy formation on the genus curve, a point emphasized by GMD and GWM.

For a Gaussian random phase distribution, the genus per unit volume is
\begin{equation}
\label{eqn_genus_expform}
g_{3D}(\nu) = A(1 - \nu^2)\exp(-\nu^2/2),
\end{equation}
where
\begin{equation}
\label{eqn_genus_expamp}
A = \frac{1}{(4\pi)^2}\frac{\int P(k)k^2d^3k}{\int
P(k)d^3k}
\end{equation}
and $P(k)$ is the power-spectrum of the smoothed density field (Doroshkevich
1970; Adler 1981; Bardeen et al.\ 1986; Hamilton, Gott \& Weinberg 1986).  The
median density contour which encloses half the volume ($\nu = 0$) is
sponge-like ($g_{3D} > 0$, i.e., many holes, see Fig.~\ref{SSDSS.3d50}\figa.
For $f < 0.16$, $\nu > 1$, we have $g < 0$, showing that we expect to see
isolated clusters [see Fig.~\ref{SSDSS.3d50}\figb].  Numerous N-body
experiments have shown that models which start off with Gaussian random phase
initial conditions, as we might expect from an inflationary model (where the
fluctuations are due to random quantum noise), retain density fluctuations with
an approximately random phase topology into the mildly non-linear regime (e.g.,
Melott, Weinberg, \& Gott 1988).  Since the biased galaxy density is generally
a monotonic function of the underlying mass density, even the biased galaxy
distribution will be approximately random phase when the contours are defined
by the volume fraction they enclose.  However, important small deviations from
random phase topology can occur.  Biasing, which systematically locates more
proto-galaxies near peaks in the initial density field, produces a small shift
in the random phase topology curve (see equation~[\ref{eqn_genus_expform}]) to
the left, a ``meatball shift'' indicating a slight preference for isolated
clusters over isolated voids.  Non-linear gravitational evolution can also lead
to a meatball shift if the smoothing length $R_s$ is of order the correlation
length.  Non-linear gravitational evolution and biasing can interact with each
other to produce an enhanced meatball effect in the standard CDM model (Park \&
Gott 1991b).  Because of these effects, the standard biased CDM $\Omega h =
0.5$, $b = 2$ model produces a slight meatball shift in the genus curve.  The
hot dark matter model, in which small scale power is damped by Landau damping,
shows a slight bubble shift (to the right) in the non-linear regime because
voids inflate to larger volume (Melott et al.\ 1988).  Whether non-linear
gravitational effects will produce a slight bubble shift or a slight meatball
shift in the topology depends on the slope of the power spectrum at
approximately the correlation length scale.  Non-linear evolution also causes
the amplitude of the genus curve to drop below that of the initial density
field (Melott et al.\ 1988; Park \& Gott 1991b; Springel et al.\ 1998).

The points in Figs.~\ref{SSDSS.3dr}\figa~and \ref{SSDSS.3dr}\figb~show genus
curves measured from a volume-limited sample of our mock SDSS catalog, in real
space and redshift space, for smoothing lengths of $R_s = 5\hMpc$ and $R_s =
10\hMpc$ respectively.  Fig.~\ref{SSDSS.3dr}\figa derives from a sample with a
depth of $R_{max} = 500\hMpc$, while Fig.~\ref{SSDSS.3dr}\figb derives from a
sample out to $R_{max} = 300\hMpc$, so that in each case the mean separation of
galaxies is approximately equal to the smoothing length.  We use a procedure
similar to that developed by Gott \etal\ (1989) for measuring the topology of
existing galaxy redshift surveys.  We first create a galaxy density field on a
grid representing a cube $1000\hMpc$ on a side, then convolve this density
field with a Gaussian smoothing filter,
\begin{equation}
\label{eqn_smoothfilter}
S(r) = {1\over{2\pi ^{3/2}R_S^3}}\exp\left({{-r^2}\over{2R_S^2}}\right),
\end{equation}
using the fast Fourier transform (FFT).  We set the smoothing length $R_s$
equal to the mean separation $\bar{d}$.  With this definition, $R_s$ is larger
by $\sqrt{2}$ than the smoothing parameter $\lambda $ used by Gott \etal\
(1989), and this smoothing criterion is therefore more conservative than that
used by Gott \etal\ (1989) and other observational analyses, which have
typically set $\lambda \approx \bar{d}$.  Because of the enormous number of
galaxies in the Sloan sample, we can afford to use a smoothing criterion that
suppresses shot noise more completely.

We set the galaxy density to zero outside the survey volume, so we must
normalize the smoothed density at a given location by the fraction of the
smoothing window that lies within the survey.  Technically, we achieve this by
creating a ``mask'' array that is one within the survey volume and zero
outside, smoothing this mask, and dividing the smoothed galaxy density field by
the smoothed mask (see Melott \& Dominik 1993).  The effective smoothing volume
near the survey boundary is thus smaller than the smoothing volume in the bulk
of the survey by up to a factor of two.  In order to avoid any systematic
biases arising in these border cells, we measure the topology only in the
volume within which the smoothed mask has a value of at least 0.8, implying
that at most 20\% of the smoothing window extends outside of the sample.

The dark points in Figs.~\ref{SSDSS.3dr}\figa~and \ref{SSDSS.3dr}\figb~show the
results of applying this procedure to the simulation's real-space galaxy
density field.  We compute the genus using a slightly modified version of the
program CONTOUR (Weinberg 1988), which is based on the curvature summation
algorithm proposed by GMD (for an alternative algorithm see Coles, Davies, \&
Pearson 1996).  We use 27 values of the threshold parameter $\nu$ (as defined
by volume fraction $f$ as described above), ranging from $\nu = -3.25$ to
$\nu = 3.25$ in steps of 0.25.  We reduce small scale jitter in the genus curve
by 
averaging over small ranges in $\nu$: a point at $\nu$ actually represents the
average genus of contours at $\langle g\rangle(\nu) = 0.2 \cdot \sum g(\nu +
\{-0.05,-0.025,0.,0.025,0.05\})$.

Ideally, we would estimate errors in the genus curve---and the covariance
matrix of the errors---using a large number of mock catalogs like the one
analyzed here, each drawn from an independent N-body simulation.  By the time
the Sloan survey is complete, a computationally ambitious approach like this
may be feasible.  For the present we rely on a less demanding procedure, using
variation within subsets of the sample to estimate the uncertainty in the mean
result.  We divide the survey volume into four quadrants along the symmetry
axes of the survey ellipse.  We measure the topology separately in each of
these four quadrants and compute a 1-$\sigma $ error bar at each value of $\nu$
from the 1-$\sigma $ dispersion of the genus in each of the four quadrants
divided by $\sqrt{3}$.  (The distribution of errors is in this case expected to
follow a Student's $t$-distribution for $N = 4$, cf. Colley [1997]).

The results obtained from the mock catalog agree with the results from the full
cube at about the level expected from the 1-$\sigma $ error bars, indicating
that our techniques for dealing with the finite survey volume do not introduce
systematic biases and that the quadrant procedure yields reasonable error
estimates.

The open points in Figs.~\ref{SSDSS.3dr}\figa~and \ref{SSDSS.3dr}\figb~show
the genus curve obtained from the redshift space galaxy distribution, with
error bars estimated from the dispersion of values in the four quadrants.
Peculiar velocities have only a small systematic effect on the genus curve,
raising it slightly at $\nu \gtrsim 1$ and lowering the peak slightly at $\nu
\approx 0$.  Matsubara (1996) noted that in the linear regime, peculiar
velocity distortions (Kaiser 1987) cause the genus curve in redshift space to
decrease in amplitude relative to the real space genus curve while retaining
the same theoretical form.

We observe in Figs.~\ref{SSDSS.3dr}\figa~and \ref{SSDSS.3dr}\figb~that
the genus curve shows no discernible left-right shift (the peak is still at
$\nu = 0$), but there are slightly more clusters observed at $\nu = 1$ than
voids at $\nu = -1$.  This effect of non-linear gravitational evolution was
first pointed out by Park and Gott (1991b), who noted its presence in the genus
curves of evolved CDM mass distributions.  They also noted that the effect was
more pronounced in the biased particle distributions because of the additional
impact of biasing.

Matsubara (1994) has calculated the expected behavior of the genus of Gaussian
random phase initial conditions after weakly non-linear gravitational
evolution.  He provides a perturbative analysis of the genus, in which odd
terms in $\nu$ add to the usual, even ($1-\nu^2$) term in the three-dimensional
genus curve for Gaussian random phase fields.  Note that Matsubara (1994) uses
the definition $\nusig = \delta/\sigma$ instead of the implicit definition in
terms of volume fraction that we adopt (eq.~[\ref{eqn_gen_nudef}]).  The two
definitions are identical only when the one-point probability distribution is
exactly Gaussian.  Matsubara's relevant result for the three-dimensional genus
is the equation
\begin{equation}
G_{3D}(\nu_\sigma) = Ae^{-\nu_\sigma^2/2}\left\{-H_2(\nu_\sigma) + \sigma\left[
{S\over 6}H_5(\nu_\sigma) + 
{{3T}\over 2}H_3(\nu_\sigma) + 
3UH_1(\nu_\sigma)\right] + O(\sigma^2)\right\},
\label{SSDSS.matseq1}
\end{equation}
where we have introduced a number of new terms.  The $H_n$'s are Hermite
polynomials of order $n$ ($H_1 = \nu_\sigma$, $H_2 = \nu_\sigma^2-1$, $H_3 =
\nu_\sigma^3 - 3\nu_\sigma$, $H_5 = \nu_\sigma^5-10\nu_\sigma^3+15\nu_\sigma$),
and $\sigma$ is the r.m.s. fluctuation in the smoothed galaxy distribution.
$S, T$ and $U$ are tabulated by Matsubara (1994) for various power spectra, the
most relevant of which is $n = -1$.

In Fig.~\ref{matsu}, we have plotted the genus in terms of both our usual $\nu$
definition for density thresholds, and in terms of strict standard deviations
($\nusig$), as Matsubara (1994) recommends.  Also, we have restricted
the range of the plot to reflect the limits suggested by Matsubara \& Suto
(1996), $-0.2 \leq \nu\sigma \leq 0.4$.  The heavy solid curve shows the random
phase theoretical form, $g_{3D} = -AH_2(\nu)e^{-\nu^2/2} =
A(1-\nu^2)e^{-\nu^2/2}$, that best fits the data points obtained from the SSDSS
(in real space), with the amplitude $A$ treated as a free parameter.  Springel
et al.\ (1998) suggest using the amplitude drop, the ratio of the fitted value
of $A$ to the value computed from the measured power spectrum via
equation~(\ref{eqn_genus_expamp}), as a diagnostic for the degree of non-linear
gravitational evolution, and Canavezes et al.\ (1998) exploit this method to
good effect in their analysis of the PSCZ redshift survey.  We have not
incorporated this technique into our present analysis, but it will certainly be
valuable for the analysis of the real SDSS.

For our fits of the pure Gaussian form of the genus curve, the $\chi^2$ values
(calculated assuming a diagonal covariance matrix with 6 degrees of freedom)
are 33 for $\nu$ and 102 for $\nusig$.  If we allow ourselves to fit
for the odd terms ($S$, $T$ and $U$) in $\nu$, we obtain the dashed curves, for
which the $\chi^2$ values are reduced to 10 and 7, respectively.  These rather
dramatic reductions indicate significant improvement in the fits, well beyond
that expected by fitting three new independent parameters alone (in which case
$\chi^2$ should decrease by 3, for three fewer degrees of freedom).

If we do not fit for $S$, $T$, and $U$ but instead use the values implied by
Matsubara's (1994) analysis given the value of $\sigma = 0.408$ measured from
the SSDSS density field smoothed at $10\hMpc$, then we obtain the dotted line
in Fig.~ \ref{matsu}.  This line improves the fit over random-phase for the
$\nusig$ genus points (open points), but is a worse fit for our usual $\nu$
points (filled points) than the original random-phase curve (with $\chi^2$
values of 56 for $\nusig$ and 144 for $\nu$).  In Table 1, we have tabulated
our fit values for $S$, $T$ and $U$ and listed the values expected by Matsubara
(1994).  While it seems that the genus curve is distorted by non-linear
evolution in the direction predicted by Matsubara (1994), the amplitude of the
distortions in the simulation is significantly smaller.  Matsubara \& Suto
(1996) have also compared cosmological simulations to these predictions and
found somewhat better agreement in some cases, particularly when $\sigma$ is
lower than our $0.408$ value.

It is perhaps not surprising that Matsubara's perturbative treatment breaks
down when the r.m.s. fluctuation amplitude is as large as $\sigma=0.408$.  For
example, in Matsubara (1994), at large $\sigma$'s the genus becomes sponge-like
again for $\nu < -2$.  Since his theoretical genus curve shows isolated voids
for $\nu = -2$, the only way to obtain a positive genus for $\nu < -2$ is if
these isolated voids begin to look like isolated sponges at lower density
thresholds, which seems implausible.  By examining Fig. 1, of Matsubara (1994),
we have found an approximate relation, $\nu \ga -1.7 - 0.18/\sigma$, which
describes the plausible applicability of the Matsubara's treatment (for $n =
1$).

The SSDSS redshift catalog yields spectacularly precise measurements of the
genus curve at these smoothing lengths ($5\hMpc$ and $10\hMpc$).  These genus
curves clearly indicate (as they should) that the initial conditions were
random phase (which they were).  The departures from the pure Gaussian form of
the genus curve, while subtle, are detected at high statistical significance,
and they have the form predicted for non-linear evolution by Matsubara (1994)
but lower amplitude.

\section{The Topology of the Peculiar Velocity Field}

The biggest uncertainty in mapping the mass density field comes from the fact
that the observed luminosity density field does not necessarily follow the 
true mass density field.  This problem, known as ``biasing,'' prevents us 
from being able to map the mass density field directly with great confidence.

We can circumvent biasing uncertainties by considering galaxies to be tracers
of the peculiar velocity field rather than the density field.  If the
gravitational instability picture is valid, peculiar velocities uniquely
reflect the mass density field.  Bertschinger \& Dekel (1989) argue that when
adopting this approach, any type of galaxy, cluster, or massive object can
serve as a test body whose motion is driven by the true density
field. Moreover, the peculiar velocity field on a given scale responds to
fluctuations of a larger scale in the density field. Since mass density
fluctuations on large scales obey linear theory better than fluctuations on
small scales, this difference in range is to our advantage.

We can use the peculiar velocity field in our topological study of large-scale
structure. If the density field is Gaussian (random phase), then the potential
field is also Gaussian, and in the linear regime, the peculiar velocity field,
being the gradient of the Gaussian potential field, should also be Gaussian
(see e.g., Bardeen et al. 1986; Park et al.\ 1992).  The {\it radial}
peculiar velocity field (which is the only measurable component of the 3D
velocity field), with the value of the peculiar radial velocity treated as a
scalar, is also Gaussian in this case (A. J. S. Hamilton, private
communication).  Thus, we can extract information about large-scale structure
by examining the topology of the radial peculiar velocity field. For example,
if, by measuring the genus curve of the radial peculiar velocity field (i.e.,
measuring the genus of iso-velocity surfaces), we find that it is not Gaussian,
we must conclude that the density field is also not Gaussian.

The SDSS will provide high resolution spectra for approximately $10^6$
galaxies. In addition to redshifts, spectra will provide other information,
including line-widths and velocity dispersions.  These measurements can be used
to obtain redshift-independent distance estimates to galaxies using methods
such as the Tully-Fisher (TF; Tully \& Fisher 1977) relation for spiral
galaxies, the Faber-Jackson (FJ; Faber \& Jackson 1976) and $D_n\mbox{-}\sigma$
(Dressler \etal\ 1987; Lynden-Bell \etal\ 1988) relations for elliptical
galaxies, and the Brightest Cluster Galaxy (BCG; Hoessel 1980; Lauer \& Postman
1994) relation. These methods typically yield r.m.s.  distance errors of 15\%
for an individual galaxy.  The TF relation can typically be applied to SDSS
spiral galaxies only if their redshifts are $z \ga 0.05$ because at smaller
distances the 3$^{\prime\prime}$-diameter fiber aperture of the SDSS
spectrographs does not subtend enough of the galaxy to yield a correct
H$\alpha$ line width.  Even with this restriction, we anticipate that the SDSS
will obtain redshift-independent distances for nearly 250,000 galaxies (Knapp
et al.\ 1997).

We treat the simulation as a realistic data set by imbuing individual
galaxies with 15\% distance errors, creating an ``observed'' catalog
that contains true redshifts of all the galaxies (since redshifts can be
measured to high precision) and estimates of the galaxies' distances.  We 
treat this ``observed'' catalog in the same way we would treat the real SDSS
dataset.  We give the galaxies estimated radial peculiar velocities
\begin{equation}
\label{eqn_potent18}
V_r=cz - H_0d,
\end{equation}
where $d$ represents the galaxies' estimated distances (with 15\% $1\sigma$
errors) and $z$ represents the galaxies' true redshifts.  In order to create
the radial peculiar velocity field, we place the galaxies (with their estimated
values of $V_r$) at their redshift distances, since at $z \ga 0.05$ the error
due to the 15\% scatter in estimated distances greatly exceeds the few hundred
km s$^{-1}$ error due to typical galaxy peculiar velocities.  It is not to our
advantage to go out to the full depth of the survey because the further out we
go the larger the errors in the velocity field become.  We therefore limit the
velocity survey at an outer depth of $r_{max} = 300\hMpc$.

Once we have radial peculiar velocity estimates for all the galaxies in our
survey volume, we smooth the data in order to obtain the smoothed large-scale
radial velocity field.  The velocity estimates for individual galaxies have
large uncertainties that are generated by the uncertainties in the distance
measurements.  The peculiar velocities of individual galaxies are usually much
smaller than these uncertaintes, making them impossible to determine
individually.  By smoothing, however, we effectively bin the velocity data in
smoothing volumes, each containing a large number of individual data points. In
this way we can beat down the noise in the $V_r$ field (velocity errors are
divided by $\sqrt{N}$, where $N$ is the effective number of galaxies contained
in one smoothing volume).  We smooth the simulated data with a Gaussian filter
of smoothing radius $R_s = 21\hMpc$.  With this choice of smoothing length, the
expected number of galaxies per smoothing volume at the outer edge of the
velocity survey is $\sim 1100$, and the expected r.m.s. error in the smoothed
radial velocity field is $\sim 130~\kms$.  Of course, the error decreases at
smaller distances.

There is one further correction that needs to be made to the ``observed''
smoothed radial peculiar velocity map. We subtract from it a smoothed
correcting map which contains the velocity field $V^\prime_r=H_0\Delta d$
(i.e. $V_r = 0$), where $\Delta d$ values are simply different randomly
generated 15\% (1-$\sigma$) distance errors for the galaxies than those
previously used. This simple correction gets rid of systematic (Malmquist-type)
effects quite well.

As we did for the galaxy density field, we calculate the genus of iso-velocity
contours in the smoothed radial velocity field using CONTOUR and estimate
errors by breaking the survey volume into four identical (in shape and volume)
sub-volumes and computing the scatter in the four independent estimates of the
genus.

Fig.~\ref{slice.vr} shows the True and ``Observed'' smoothed radial peculiar
velocity fields (note that this is a different slice than shown in Fig.
\ref{bigslice}). From Fig.
\ref{slice.vr}, it is fairly obvious that the radial peculiar velocity field
can be mapped out to a large scale, despite the large uncertainties in our
distance measurements. The SDSS data should be sufficient for this
purpose. Fig.~\ref{slice.vr} shows that the ``Observed'' map captures the
most prominent features of the True map. There is some distortion, as expected,
but overall the maps compare reasonably. The r.m.s. error in our ``Observed''
map (i.e., the r.m.s. pixel by pixel difference between it and the True map),
$121.0~\kms$, is smaller than the actual velocity amplitudes seen in our True
map (whose r.m.s. value is $148.8~\kms$).

The genus curves for both the True and ``Observed'' (Fig.~\ref{genus.obs})
radial peculiar velocity fields are very noisy, as expected, because of the
small number of resolution elements in our survey volume.  Nevertheless, both
genus curves are roughly Gaussian (the theoretical curve fits our data points
as well as we would expect given the 1-$\sigma$ error bars).

\section{The 2-D Topology of Redshift Slices}

Before discussing the 2-D topology of redshift slices, we should 
explain the somewhat untraditional coordinate system of the SDSS.  
A galaxy's position on the sky is defined by a survey latitude $\eta$
and a survey longitude $\lambda$, but the nature of the constant
latitude and constant longitude curves is backwards from the usual;
the constant latitude curves are great circles that connect the
survey poles (an {\it east} pole at $\delta=0$, $\alpha=18^h20^m$
and a {\it west} pole at $\delta=0$, $\alpha=6^h20^m$), and the
constant longitude curves are small circles centered on these poles.
The SDSS imaging observations are carried out in scanning mode
(see Gunn et al.\ 1998), and the constant latitude curves are the scan tracks.

Since imaging must precede spectroscopy in any given area of the survey,
an early product from the SDSS is likely to be redshift survey slices,
much like those of the LCRS.  We have therefore selected six
slices of constant survey latitude from the mock redshift catalog
for 2-D topological analysis.
These slices are centered on arcs of constant $\eta$, but they have
constant angular width in the sky, not constant $\Delta\eta$; thus, while the
slices' centers are great circles, their upper and lower boundaries are not.
In order to allow more direct comparison to existing results from the LCRS, we
have made these slices $1.5\deg \times \sim 80\deg$, rather than the
$2.5\deg \times 130\deg$ of a full SDSS imaging stripe.
We choose three slices at low latitude
($\eta = -33\deg, -30\deg, -27\deg$) and three at high latitude
($\eta = 27\deg, 30\deg, 33\deg$), again to obtain a sample similar
to that of the full LCRS.

The galaxy sample in the SSDSS is, roughly speaking,
magnitude-limited at $m_{max}=17.9$ in $r^\prime$ (see Section 3).  As
with any magnitude-limited survey, conversion of counts to real galaxy density
requires a good understanding of selection effects.  We can compensate for
these effects by constructing the two-dimensional selection function, which at
a given radius reflects the expected surface density of galaxies in the
survey.

First we find the maximum distance, $D_{max,i}$, at which the $i$th galaxy
(with actual distance $D_i$ and magnitude $m_i$) could be seen.  To a good
approximation, this maximum distance is
\begin{equation}
D_{max,i} = D_i \cdot 10^{0.2(m_{max}-m_i)},
\end{equation}
though several small complications exist; we include effects such as
limiting surface brightness and {\it K}-corrections in
constructing $D_{max}$.  With each $D_{max,i}$ computed, we can invoke 
Schmidt's (1968)
$V/V_{max}$ method to construct the expected volume-density of galaxies
as a function of radius,
\begin{equation}
\rho_s(r) = {3\over{\Omega_s}} \sum_{D_{max,i}>r}{D_{max,i}^{-3}}
\end{equation}
(Gott \etal\ 1989), where $\Omega_s$ is the solid angle of the slice.  We
multiply $\rho_s$ by radius $r$ to account for the linearly expanding
wedge-shaped profile of the slice, and also by a ``shape factor'' $S$ (Park
\etal\ 1992) which relates solid angle in the sky to azimuthal angle in the
slice map
\begin{equation}
\sigma_s(r) = Sr\rho_s(r).
\end{equation}
Here $S = [2\sin(w/2)]$, where $w$ is the constant angular width of the slice.
Recall that the center of the slice follows a great circle of constant $\eta$,
which causes a difference from the shape factor given by Park \etal\ (1992)
for declination slices, which do not follow great circles.

We now cut a slice from this azimuthally symmetric selection function with the
same longitude spread as the actual slice.  We then smooth both the selection
function slice and the actual survey slice with a Gaussian disc,
$e^{-r^2/2R_s^2}$.  When we divide the smoothed survey slice by the
smoothed selection slice, we have a map where the effects of the flux limit
(and edge effects) have been minimized.  We truncate the slice at the radius
where the smoothing length, $R_s$, equals $\sigma_s^{-1/2}$ to avoid shot
noise sampling effects (we actually truncate one smoothing length inside of
that radius to reduce these effects further).

We have chosen our smoothing length to ensure that most of the structures
detected are well within the linear regime, because in the linear regime
fluctuations at fixed position simply grow in amplitude in proportion
to the growth factor, so that the topology of the
present-day fluctuations is similar to the topology in the initial
fluctuations.  Non-linear effects typically become important on scales below
about $8\hMpc$, so we have chosen a smoothing convolution kernel of $\exp(-r^2 /
2R_s^2)$, with $R_s = 20\hMpc$.  (As mentioned previously, 
Matsubara [1996] has shown that peculiar velocities in the linear regime
do not distort the form of the genus curve expected for Gaussian fluctuations.)

Fig.~\ref{SSDSS.slice} provides a contour map of one of the smoothed,
calibrated slices ($\eta = -30\deg$, width $1.5\deg$), with the galaxy
locations over-plotted.  The heavy lines are contours of high density; the
lighter lines are contours of low density, and the dashed line is the median
density contour.  We will discuss the exact values of these contours below;
for now we just wish to illustrate that the contours described
correctly identify real over-densities and voids in the data.  Also, many
``fingers of God'' are visible in the data; these are indicators of non-linear
effects on small scales (such as virialized clusters). The structures visible
in the contours, however, are much larger than the fingers of God, which
suggests that we have smoothed out most of the non-linear features in the data
(the r.m.s. fluctuation amplitude in the slices is
$\sigma_{20h^{-1}\mbox{\scriptsize Mpc}}$ = 0.3).

As with the LCRS, we can immediately note the large number of structures,
critical to a quantitative analysis via the genus statistic.  When compared
with Park \etal\ (1992), which uses the Geller \& Huchra (1989) survey, we find
a vast improvement in the number of structures detected (by about a factor of
10), thus a much stronger lever-arm with which to measure topology statistics.

Following Melott \etal\ (1989) and Gott \etal\ (1990), we define the
two-dimensional genus $G_{2D}$ of the excursion set for a random density field
on a plane as
\begin{equation}
\begin{array}{ll}
G_{2D} = &\mbox{(number of isolated high-density regions)}- \\
~&   \mbox{(number of isolated low-density regions)}.\\
\end{array}
\end{equation}
Equivalently, the genus can be defined as the total curvature of the contours.
Assuming a contour defines a differentiable curve $C$ on the map, its total
curvature is given by the integral
\begin{equation}
K = \int_C \kappa ds \equiv 2\pi G_{2D},
\end{equation}
where $\kappa$ is the local curvature, $s$ parameterizes the curve, and $G$ is
the genus of the contour.
An isolated overdense region will contribute $+1$ to the total map genus and a
void (``hole'') in it will decrease the genus by 1.  In practice, contours may
cross the edge of the survey region; in that case the partial curves contribute
non-integer rotation indices to the genus.

A two-dimensional random phase Gaussian density field will generate a genus per
unit area
\begin{equation}
g = A \nu e^{-\nu^2/2},
\label{SSDSS.eqgen2d}
\end{equation}
where $A$ is a constant and $\nu$ is the threshold value, 
related to the area fraction $f$ by equation~(\ref{eqn_gen_nudef}).
The value of $A$ depends on an integral of the detailed power spectrum of the
fluctuations, but in the case of a perfect power-law spectrum with index $n >
-1$,
\begin{equation}
A_{n>-1} = {1\over{2\cdot(2\pi)^{3/2}R_s^2}},
\end{equation}
where $R_s$ is the Gaussian smoothing length (Melott \etal\ 1989;
see also Adler 1981; Coles 1988; Park \etal\ 1990; Gott \etal\ 1992).  We have
explicitly used the area fraction in order to be less sensitive to non-linear
effects and biasing as discussed in Section~4.
Simulations have shown that the genus curve defined in this way more
reliably reflects the genus curve of the initial fluctuations,
whose nature we wish to test.

In Fig.~\ref{SSDSS.slice}, we have plotted the contour map of the smoothed
density distributions in one of the SSDSS slices, with contours at $\nu =
\{-2,-1,0,1,2\}$.  The heavy lines represent $\nu = \{1,2\}$, the light lines
$\nu = \{-2,-1\}$, and the dashed line $\nu = 0$.  The map shows many
excursions at non-zero values of $\nu$, while the median contour ($\nu = 0$)
wanders through the map rather randomly, as expected.

In Fig.~\ref{SSDSS.2dgen}, we have plotted as a function of $\nu$ the mean
genus per unit area (averaged from the estimates in each of the six slices).
The best-fit theoretical genus curve expected for a random phase Gaussian
distribution, equation~(\ref{SSDSS.eqgen2d}),
is shown as a solid curve,
with a best-fit value of $A = 0.7A_{n>{-1}}$.  The errorbars are the 68\%
(solid) and 95\% (dotted) confidence limits, estimated from the formal
Student's $t$-distribution for $n = 6$ (six slices) (Lupton 1993).  This
somewhat less familiar distribution is necessary whenever the error is
estimated from the data distribution directly, as opposed to an independent
estimation of the error (see Colley 1997).  The $t$-distribution is 
equivalent to a Cauchy (Lorentzian) distribution for $n = 2$, 
but it converges to a Gaussian
distribution quite rapidly (by $n = 20$, the difference from Gaussian is
negligible in most applications).  The $t$-distribution has broad wings to
allow for accidentally low sample variances ($s^2$), which cause the
$t$-variate, $(\bar{x} - \mu) / (s/\sqrt{n-1})$, to reach anomalously
high values relative to the normal variate, $(\bar{x} - \mu) /
(\sigma/\sqrt{n})$, ($\bar{x}$ is the sample mean, $\mu$ is the true mean, and
$\sigma$ is the true standard deviation).

The Gaussian-field theoretical curve (solid line) in 
Fig.~\ref{SSDSS.2dgen} fits the
SSDSS data points reasonably well.  To be more quantitative
about this, we perform a test related to the $\chi^2$ statistic.  Our
``not-quite $\chi^2$ statistic,'' $\tilde{\chi}^2$, is computed in the usual
way, but is different from a formal $\chi^2$ in that we have used the formal
1-$\sigma$ errors as estimated from the six slices (recall the genus
measurements are $t$-distributed variates, not Gaussian),
\begin{equation}
\tilde{\chi}^2 = \sum_{i = 1}^{21}
{{(\bar{g}_i-\tilde{g}_i)^2}\over{\sigma_{\bar{g}_i,est}^2}} .
\end{equation}
Here, the sum runs over the 21 values of $\nu$ where we have measured the genus
(as shown in Fig.~\ref{SSDSS.2dgen}); $\bar{g}_i$ is the mean genus among the
six slices at each $\nu$-value; $\tilde{g}_i$ is the value of the fitted solid
curve, and $\sigma_{\bar{g}_i,est}$ is the formal standard error in the mean as
estimated from the six slices.  For comparison, we ran $10^4$ simulations of 20
$t$-variates with $n = 6$ (one less than 21, due to the one-parameter fit).  We
then computed $\tilde{\chi}^2$ for these datasets, and found that our observed
value of 36.4 fell at the 70\% confidence level (i.e., one would expect the
value of $\tilde{\chi}^2$ to be less than ours 70\% of the time, more 30\% of
the time).  Thus, the results were nearly within a 1-$\sigma$ statistical
agreement with the fitted curve theoretical curve $g(\nu) \propto \nu
e^{-\nu^2/2}$.

Although the above test accounts for the non-Gaussian error distribution
on individual points, it does not account for the covariance of
errors from one value of $\nu$ to another.
As one can see from Fig.~\ref{SSDSS.2dgen}, 
the difference between the solid curve and the data points is systematic
and coherent: the best-fit curve underestimates $|g(\nu)|$ at $\nu>0$
but overestimates $|g(\nu)|$ at $\nu<0$.  In order to assess the
significance of this systematic departure from the Gaussian-field
prediction, we turn to a more elaborate statistical technique
developed by Colley (1997).  Alternative approaches to dealing with
correlated errors in genus curves are described by Vogeley et al.\ (1994),
Protogeros \& Weinberg (1997), and Springel et al.\ (1998).

Since the genus values at nearby $\nu$ values might be correlated (they are
measuring essentially the same structures), we should look for a point-to-point
correlation among the measured genus values.  If the correlation is
significant, there will be significant off-diagonal terms in the covariance
matrix formed from the genus among the various $\nu$ values.  In this case, the
commonly used $\chi^2$ treatment, which assumes a diagonal covariance matrix,
must be replaced by a treatment which explicitly employs the full covariance
matrix in the $\chi^2$ computation.  This more formal $\chi^2$ statistic should
be an apt figure of merit for the quality of fit of the theoretical genus curve
to the genus values derived from a two-dimensional density field.

Since the point-to-point covariance is not known a priori, we are left to
produce independent model fields to estimate the covariance.  The model fields
are pure Gaussian random phase fields, where each Fourier mode has a random
phase and amplitude derived from a Gaussian distribution with a mean of zero,
and a variance equal to the value of the power spectrum $P(|\vec{k}|)$ at the
$\vec{k}$ position of the mode in Fourier space.  We used the two-dimensional
power spectrum for an $\Omega = 0.4, \Omega_\Lambda = 0.6, h = 0.6$ cosmology
(i.e. the same power spectrum as in the simulation) in generating 100 sets of 6
Gaussian random phase slices with identical physical dimensions to the SSDSS
slices.  We applied exactly the same smoothing and genus topology routines to
each model field as we did to the SSDSS slices, giving us a total of 101
independent model genus datasets, 100 of which are guaranteed to derive from
Gaussian, random phase fields.  For each model dataset, $m$, we computed the
best fit theoretical curve, and derived its pairwise covariance between the
various values of $\nu_i$ and $\nu_j$.  Model $m$ has covariance
\begin{equation}
C_{ij,m} = (\bar{g}_{i,m} - \tilde{g}_{i,m})(\bar{g}_{j,m} -
\tilde{g}_{j,m}).
\label{SSDSS.eqcovmat}
\end{equation}
Treating all datasets equivalently, we leave out each one in turn, and compute
from the remaining 100 a model covariance matrix as the average of $C_{ij,m}$
over the remaining $m$ values, \ie\ $\langle C_m\rangle_{ij} = \sum_{m^\prime
\neq m}{C_{ij,m^\prime}}/100$.  This allows a direct computation of $\chi_m^2$
in dataset $m$, based on a completely independent covariance matrix.
\begin{equation}
\chi_m^2 = \sum_{i,j} (\bar{g}_{i,m} -
\tilde{g}_{i,m})\langle C_m\rangle_{ij}^{-1}(\bar{g}_{j,m} - \tilde{g}_{j,m}).
\label{SSDSS.eqcovchi2}
\end{equation}
With this improved figure of merit for the fit, we can compare the fit derived
from the observed data with that derived from the 100 model datasets.  Without
further assumption we may evaluate directly the rank of $\chi^2$ from the real
dataset among the 101 $\chi^2_m$ values.  We obtain for the SSDSS dataset a
rank of 97 out of 101.  The probability of obtaining a result that dramatic by
chance is 5 out of 101.  The simulated slices are therefore just marginally
inconsistent with the pure Gaussian random phase fields at the 95\% confidence
level.  Colley (1997), using identical techniques, found that the LCRS agreed
better with the simple Gaussian random phase genus curve, ranking 67th out of
101 in its $\chi^2$ when compared to pure Gaussian random phase curves.  The
small but significant departure of the SSDSS from the Gaussian random phase
curve is presumably due to non-linear effects of structure growth, since the
SSDSS was seeded with Gaussian random phase initial conditions.  In the
following, we discuss non-linear effects on the genus and assess the
statistical distinguishability of the LCRS genus and SSDSS genus.

We reconsider Matsubara's (1994) equation for the 3-D genus curve after weakly
non-linear evolution of structure, equation (\ref{SSDSS.matseq1}).  In dropping
from three to two dimensions, we would expect the Hermite polynomial indices to
drop by one, although the coefficients would not necessarily change in a
trivial way.  The treatment above fits for the coefficient of
$H_1(\nu) = \nu$, the only term expected in a purely Gaussian random phase
field (eq.~[\ref{SSDSS.eqgen2d}]).  This fit is shown by the solid
curve in Fig.~\ref{SSDSS.2dgen}.  If we now allow fitting for even terms,
($H_0 = 1$, $H_2 = \nu^2-1$, $H_4 = \nu^4 - 6\nu^2 + 3$) we find a much better
fit, shown by the dashed curve in Fig.~\ref{SSDSS.2dgen}.  Adding these terms
reduces the number of voids by 8\% ($\nu = -1$) and increases the number of
clusters by 16\% ($\nu = 1$) relative to the best fit with equal numbers of
clusters and voids.  As in the three-dimensional case, the new terms
drastically reduce the $\chi^2$ of the fit, from 36.4 to 13.1 (even a bit
below the expected level of 17), much more of a reduction than 3, as expected
when adding three new independent adjustable parameters.  We also note that the
apparent effects of non-linearity work in the same sign in both the
three-dimensional and two-dimensional cases, in that there are slightly more
clusters than voids in both cases.  The coefficients of the even terms are
roughly one order of magnitude smaller than the $H_1$ coefficient, which again
suggests that the effects of non-linearity seem to work in the direction
suggested by Matsubara (1994), but not as substantially.

For comparison with real observations, we have included an identical plot to
Fig.~\ref{SSDSS.2dgen} for the LCRS (Fig.~\ref{SSDSS.lcrsgen}).  Upon
inspection, one sees that, as in the SSDSS, the number of clusters in the LCRS
is larger than the number of voids, but the effect is not as large as it is in
the SSDSS.  Using identical statistical techniques to those above, Colley
(1997) showed that the LCRS genus curve was consistent with a Gaussian random
phase genus curve within the 1-$\sigma$ confidence limit, while here we have
shown that the SSDSS was marginally inconsistent at the 2-$\sigma$ (95\%)
level.  In fitting for the non-linear terms in the LCRS, we have found that the
coefficients of $H_0$, $H_2$ and $H_4$ imply a 6\% decrease in voids at $\nu =
-1$, and a 7\% increase in clusters at $\nu = 1$, relative to the best fit
curve with equal numbers.  When compared to the 8\% decrease in voids and 16\%
increase in clusters in the SSDSS, we see that the LCRS genus curve is less
distinguishable from random phase than is the SSDSS genus curve.  Furthermore
adding the three new fit parameters to the LCRS genus curve decreased $\chi^2$
by only 4, (roughly) as expected when adding three new independent parameters.
In the case of the SSDSS, adding the three new parameters decreased $\chi^2$ by
23, indicating that the new parameters were important to the fit.

A more direct comparison of the LCRS and SSDSS reveals that the genus curves
derived from these surveys are consistent with each other.  Using the
$\tilde\chi^2$ method described in the previous section, we have found that the
LCRS and SSDSS are consistent well within the 1-$\sigma$ consistency criterion.
We now use a covariance matrix method similar to that discussed in the previous
section (equations [\ref{SSDSS.eqcovmat}] and [\ref{SSDSS.eqcovchi2}]); however
instead of using $(\bar{g}_{i,m} - \tilde{g}_{i,m})$, the sample mean minus the
best-fit value, as the deviate at the $i$th $\nu$ value, we use
$(\bar{g}_{S,i,m} - \bar{g}_{L,i,\ell})$, the mean genus value in SSDSS data
set $m$ minus the mean genus value in LCRS dataset $\ell$, where each dataset
includes six slices.  As before, we generated 100 fake datasets with six pure
Gaussian random phase slices with identical geometry, power-spectrum and
smoothing to those of the ``real'' datasets, for both the SSDSS and LCRS.  This
means we have 101 independent datasets for both the SSDSS and LCRS.  We then
use equations (\ref{SSDSS.eqcovmat}) and (\ref{SSDSS.eqcovchi2}), with the
above substitutions to compute $\chi_{\ell m}^2$ for the 10201 possible
combinations of datasets:
\begin{equation}
\langle C_{\ell m}\rangle_{ij} = (10200)^{-1}\sum_{\ell^\prime,m^\prime \neq
\ell,m} (\bar{g}_{S,i,m^\prime} - \bar{g}_{L,i,\ell^\prime})
(\bar{g}_{S,j,m^\prime} - \bar{g}_{L,j,\ell^\prime}),
\end{equation}
\begin{equation}
\chi^2_{\ell m} = \sum_{i,j = 1}^{21} (\bar{g}_{S,i,m} -
\bar{g}_{L,i,\ell}) \langle C_{\ell m}\rangle^{-1}_{ij}
(\bar{g}_{S,j,m} - \bar{g}_{L,j,\ell}).
\end{equation}
We find that $\chi^2$ for the $\ell = m = 101$ dataset, where the real datasets
of both the LCRS and SSDSS are compared, falls at the $46th$ percentile of all
10201 $\chi^2$ values, an excellent agreement.  This consistency between the
LCRS and SSDSS is quite remarkable, as is recognizable when comparing Figs.
\ref{SSDSS.2dgen} and \ref{SSDSS.lcrsgen} directly.  In fact, the statistical
similarity extends to the best-fit the amplitudes of the genus curves:
$A_{LCRS} = 53$ and $A_{SSDSS} = 56$ (see equation [\ref{SSDSS.eqgen2d}]).  The
expected error in each amplitude (known from the simulated sets) is of order 3.
The consistency between these two datasets indicates that the model cosmology
within the simulation produces a remarkably similar genus curve to that derived
from the observations.

\section{Availability of the Mock Catalog}

We anticipate making further use of this mock redshift catalog in
our own preparations for analysis of the SDSS.  In the hope that
it may be useful to other researchers both inside and outside the
SDSS collaboration, we are making the mock catalog available at
{\tt http://www.astronomy.ohio-state.edu/$\sim$dhw/ssdss.html}
(in the event of questions or difficulties, contact David Weinberg).
This catalog complements the set of publicly available
2dF and SDSS mock catalogs described by Cole et al.\ (1998).
The present catalog is drawn from a larger volume simulation
[$(600\hMpc)^3$ vs. $(345.6\hMpc)^3$] and uses more careful modeling
of the SDSS selection criteria.  The Cole et al. simulations, on
the other hand, have higher gravitational force resolution 
($90h^{-1}$ kpc vs. $1\hMpc$) and cover many cosmological models
(20 different sets of cosmological parameters and several biasing schemes).
Thus, the present mock catalog is probably more suitable for studies
that probe large scales or require careful matching of the 
anticipated SDSS selection function, while the Cole et al.\ mock catalogs
are more useful for studies aimed at testing the ability of statistical
diagnostics to distinguish between cosmological models with 
surveys the size of 2dF or the SDSS.  For those wishing to create
mock SDSS catalogs from their own large N-body simulations, the code
used to assign galaxy properties and apply the SDSS selection criteria
is available on request from David Weinberg.  It is not trivial
to use, but it is extensively commented.

\section{Conclusions}

We have constructed a mock survey to mimic the SDSS redshift survey of a
million galaxies in the North Galactic Cap.  We have used a large $N$ =
54,872,000 body simulation with $\Omega_{CDM} = 0.4$, $\Omega_\Lambda = 0.6$,
$h = 0.6$, $b = 1.3$, which has the Gaussian random phase initial conditions
expected from an inflationary model, where perturbations arise from quantum
fluctuations in the early universe.  We measure the 3-D genus curve in this
simulation and find that it indeed exhibits the shape predicted theoretically
for Gaussian initial conditions.  The SDSS will be able to measure the genus
curve with unprecedented precision.  A sponge-like topology with $\sim 500$
holes is measured here, giving a point-wise statistical precision of 4\%, a
vast improvement over previous surveys.  Indeed, the data trace the random
phase curve so well that observers confronted with such data would almost
certainly conclude (correctly) that the mock universe had started with Gaussian
random phase initial conditions.  Small deviations from the random phase curve
are, however, detectable at high statistical significance, the main effect
being a slight excess of clusters over voids that results from non-linear
gravitational evolution and biasing (Park \& Gott 1991b).  Although the
influence of non-Gaussanity of primordial fluctuations on the genus curve
dependes on the specifics of the theoretical model, the numerical studies of
the topolgy of the texture model by Gooding \etal\ (1992) and of ``generic''
non-Gaussian models by Weinberg \& Cole (1992) suggest that typical
non-Guassian models produce distortions of the genus curve that could be
detected easily at the high level of precision obtainable with the SDSSS.

In comparing to predicted effects of non-linear structure growth on the genus
(Matsubara 1994), we find the measured effects to be consistent in sign
but not in magnitude.  The Matsubara (1994) formula predicts a much larger
(roughly two orders of magnitude) distortion of the genus curve than we
observe.  This discrepancy may result from using the 
formula outside its applicable range.

We have also shown how the genus curve of the velocity field may be measured
in the SDSS.  Again, the results are consistent with the Gaussian random phase
prediction, but the statistical precision is very low because
a very large smoothing length is required to beat down the noise
in redshift-independent distance measurements.

Since the earliest products from the Sloan redshift survey will be 2-D slices,
we have measured the topology (genus) of large-scale structure in such slices
drawn from the mock survey.  Again, we find that the number of clusters
slightly exceeds the number of voids, but at a lesser level than would be
presumably expected from Matsubara (1994).

Finally, we compare our simulation slices directly with a very similar study of
the (observed) Las Campanas Redshift Survey slices (Colley 1997).  While both
the LCRS and our simulation slices show a slight excess of clusters over voids,
the effect is about twice as large in the simulation.  The genus of the LCRS
slices nonetheless agrees with the genus of the simulation slices, well within
the 1-$\sigma$ interval, without any adjustment of amplitude in the curves.

A better understanding of the non-linear structure growth and its effects on
the genus curve will come with ever more elaborate cosmological simulations,
and with the Sloan Digital Sky Survey itself.  A survey this large is a
powerful tool for evaluating theories of cosmic structure formation.

This paper is supported by NSF grant AST95-29120 and NASA grant NAG5-2759. We
thank Bharat Ratra, Andrew Hamilton, Martin Bucher and Neta Bahcall for helpful
conversations.  We thank Michael Strauss for computing the K-corrections
and galaxy colors used in the mock catalog.
DW thanks Jim Gunn, Hans-Walter Rix, and Michael Strauss for numerous
helpful discussions on the assignment of galaxy properties, and he
acknowledges the support of a Keck fellowship at the Institute for Advanced
Study during the early phases of this work.
CP acknowledges the support of the Basic Science Research
Institute Program, Ministry of Education 1995 (BSRI-98-5408).

\vfill
\newpage
\begin{table}[p]
\begin{center}
{\bf Table 1}
\vskip 1cm
\begin{tabular}{cccc}
\hline
\hline
&predicted &best-fit ($\nu$) &best-fit ($\nusig$) \\
S & 3.468 & $-0.136$  & $ 0.0501$\\
T & 2.312 & $-0.0314$ & $-0.116$ \\
U & 1.227 & $ 0.129$  & $ 0.0512$\\
\hline
\hline
\end{tabular}
\end{center}
\caption{Best-fit values of $S$, $T$ and $U$ vs. values predicted by Matsubara
(1994).  In describing the genus of the large-scale distribution of galaxies,
$S$, $T$ and $U$, defined in equation (\ref{SSDSS.matseq1}), are coefficients
of odd polynomials which add to the usual 2nd-order even polynomial, expected
in the Gaussian random phase genus curve of three dimensions.  The second
column is for density thresholds described by $\nu$ (equation
\ref{eqn_gen_nudef}); the third column is for density thresholds defined
strictly by standard deviations $\nusig = \delta/\sigma$.  Note that the range
of the density thresholds for the fits has been limited to the range suggested
by Matsubara \& Suto (1996), who give $-0.2 \leq \nu\sigma \leq 0.4$, where
$\sigma$ is the standard deviation in density fluctuation at this smoothing
length ($R_s = 10\hMpc, \sigma = 0.408$).}
\end{table}
%

\vfill

\pagebreak
\begin{figure}[p]
\caption{A projection onto the sky of the nearly 1 million galaxies in
the Simulated Sloan Digital Sky Survey.  The tickmarks indicate the boundaries
of the six degree slice in Fig.~\ref{bigslice}.}
\label{skyfig}
\end{figure}

\begin{figure}[p]
\caption{A six degree slice of the Simulated Sloan Digital Sky Survey}
\label{bigslice}
\end{figure}

\begin{figure}[p]
\caption{\figa~Median density contour in the Simulated Sloan Digital Sky
Survey, showing its sponge-like form.  The observer is at the apex (left) and
the radius of the sample is $500\hMpc$, and the smoothing length (\cf\ equation
[\ref{eqn_smoothfilter}]) is $R_s = 10\hMpc$. }
\label{SSDSS.3d50}
\end{figure}

\addtocounter{figure}{-1}
\begin{figure}[p]
\caption{\figb~93rd density percentile contour in the Simulated Sloan
Digital Sky Survey, showing isolated clusters, for the same sample 
and smoothing length shown in Figure
\ref{SSDSS.3d50}\figa.}
\label{SSDSS.3d93}
\end{figure}

\begin{figure}[p]
\plotfiddle{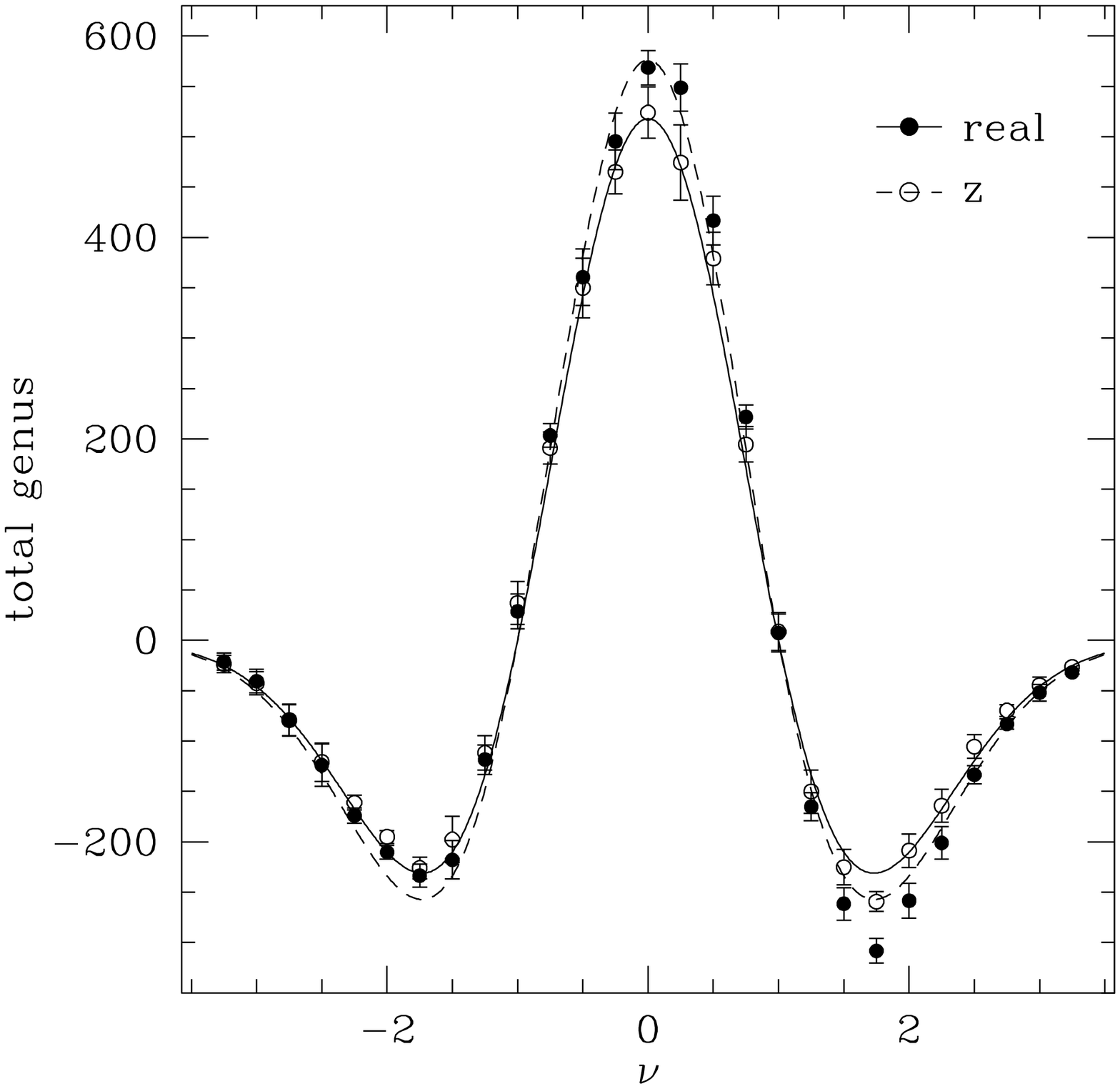}{13cm}{0}{80}{80}{-240}{-110}
\caption{\figa~The genus curve of three-dimensional structure in the SSDSS,
in real space (filled points, and solid curve), and redshift space (open points
and dashed curve).  The smoothing length is $R_s=5\hMpc$.  The curves are the
best fit for $g = A\cdot(1-\nu^2)\exp(-\nu^2/2)$, expected for a Gaussian
random phase field.  Errorbars are 1-$\sigma$ confidence intervals.}
\label{SSDSS.3dr}
\end{figure}

\addtocounter{figure}{-1}
\begin{figure}[p]
\plotfiddle{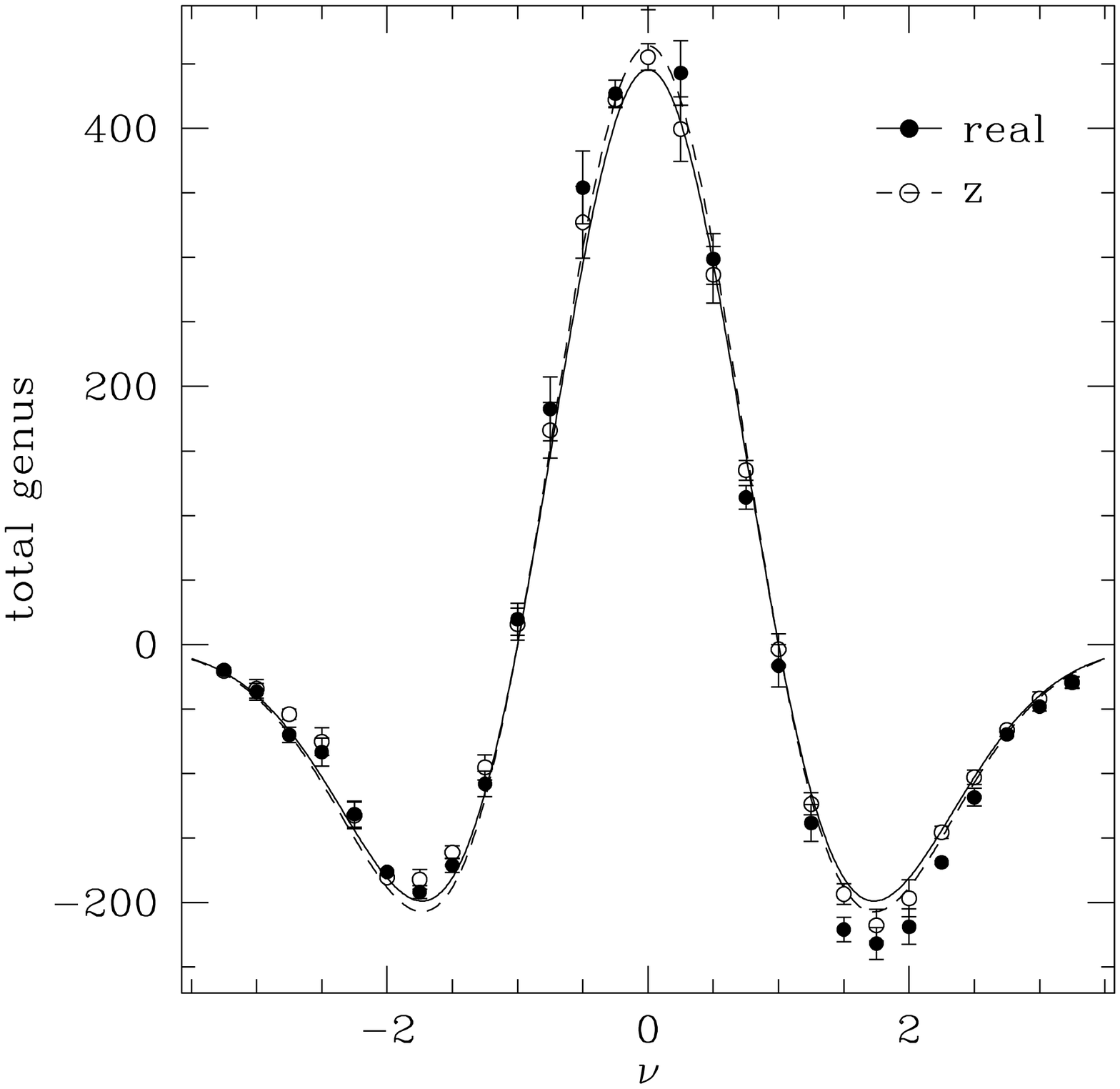}{13cm}{0}{80}{80}{-240}{-110}
\caption{\figb~as in \ref{SSDSS.3dr}\figa, but for a smoothing length
of $R_s = 10\hMpc$.}
\label{SSDSS.3dz}
\end{figure}

\begin{figure}[p]
\plotfiddle{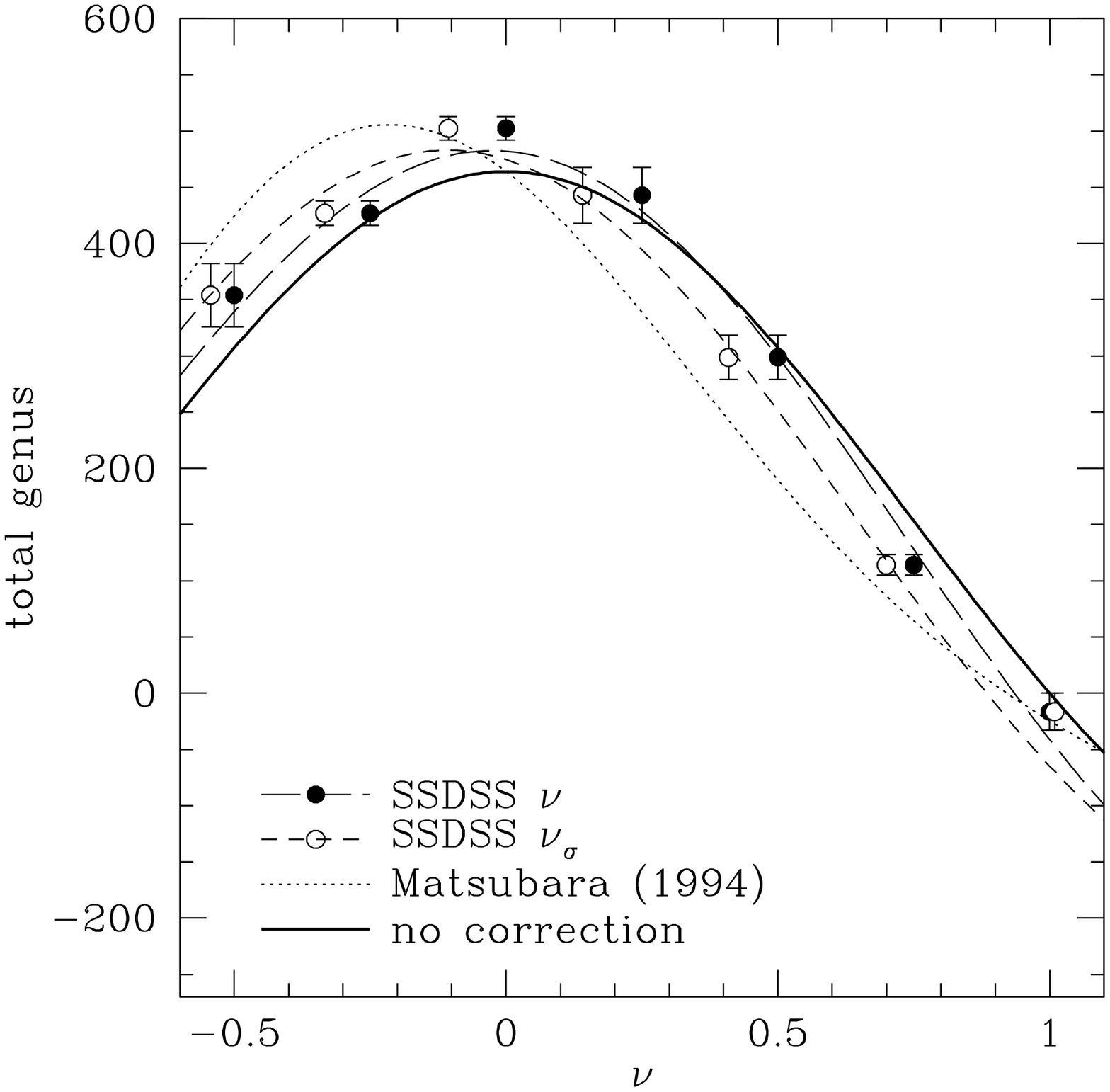}{13cm}{0}{80}{80}{-240}{-110}
\caption{Comparison of the 3-D genus with predictions from Matsubara (1994) for
weakly non-linear evolution of the genus.  The heavy solid curve is the best
fit for the purely Gaussian random phase curve.  The solid points reflect
density thresholds $\nu$ defined in equation \ref{eqn_gen_nudef}; the open
points reflect density thresholds which are strict standard deviations $\nusig
= \delta/\sigma$.  The long- and short-dashed curves include the best fits for
amplitude and for $S$, $T$ and $U$, given in equation (\ref{SSDSS.matseq1}), in
terms of $\nu$ and $\nusig$ respectively.  These are coefficients of odd terms
added to the usual even genus curve. The dotted curve is the prediction of
Matsubara (1994).  Note that the range of the density thresholds is delimited
by Matsubara \& Suto (1996), who suggest $-0.2 \leq \nu\sigma \leq 0.4$, where
$\sigma$ is the standard deviation in density fluctuation at this smoothing
length ($R_s = 10\hMpc; \sigma = 0.408$).}
\label{matsu}
\end{figure}

\begin{figure}[p]
\caption{At top is the true radial peculiar velocity field in a slice of the
simulated SDSS. At bottom is the ``observed'' radial peculiar velocity 
field.  Both fields have been smoothed with a Gaussian filter of radius 
$R_s = 21\hMpc$.  The shaded regions have a positive radial peculiar velocity
and the shaded contours represent velocities of $+50~\kms$ and $+100~\kms$.
The unshaded regions have a negative radial peculiar velocity
and the unshaded contours represent velocities of $-50~\kms$ and $-100~\kms$.}
\label{slice.vr}
\end{figure}

\begin{figure}[p]
\plotfiddle{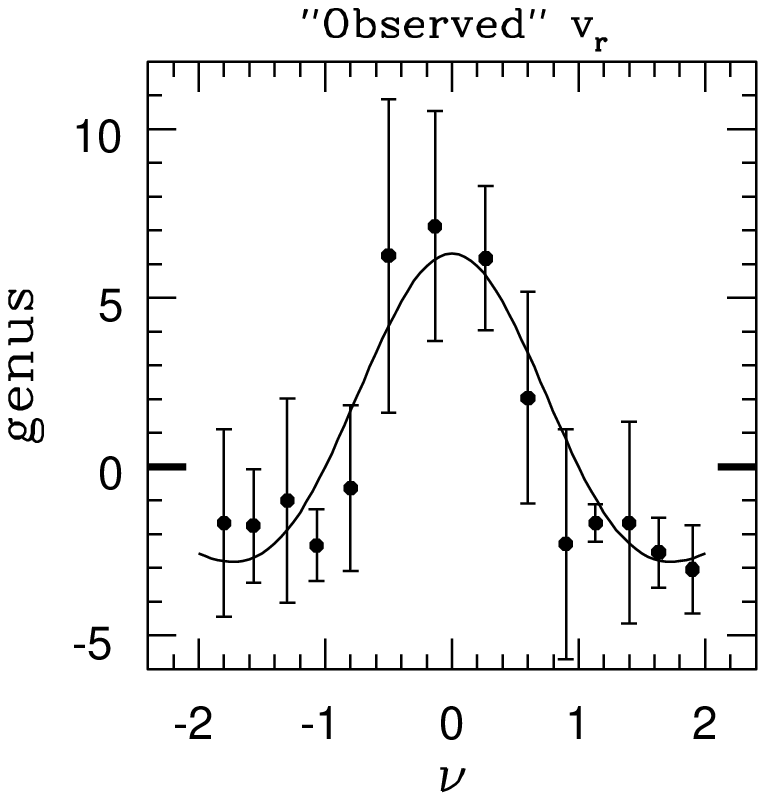}{16cm}{0}{100}{100}{-420}{-220}
\plotfiddle{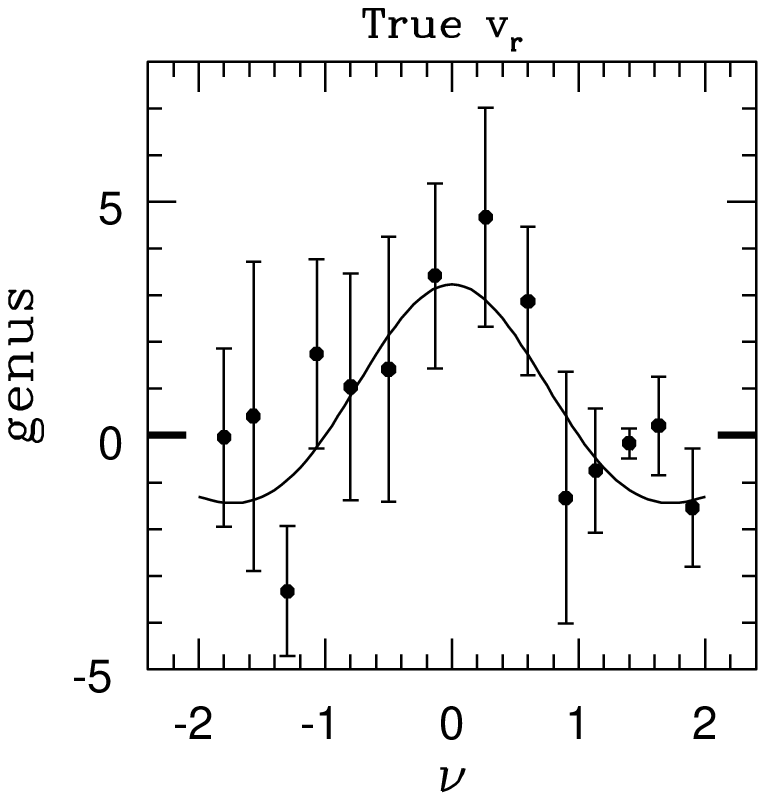}{0cm}{0}{100}{100}{-175}{-188}
\caption{The genus curve of the ``observed'' radial peculiar 
velocity field (left) and the true radial 
peculiar velocity field (right) of the SSDSS.
Points show simulation measurements with 1-$\sigma$ error bars
computed from the variance among four subsets of the survey,
and smooth curves show the form expected for a Gaussian field
(eq.~[\ref{eqn_genus_expform}]).
}
\label{genus.obs}
\end{figure}

\begin{figure}[p]
\plotfiddle{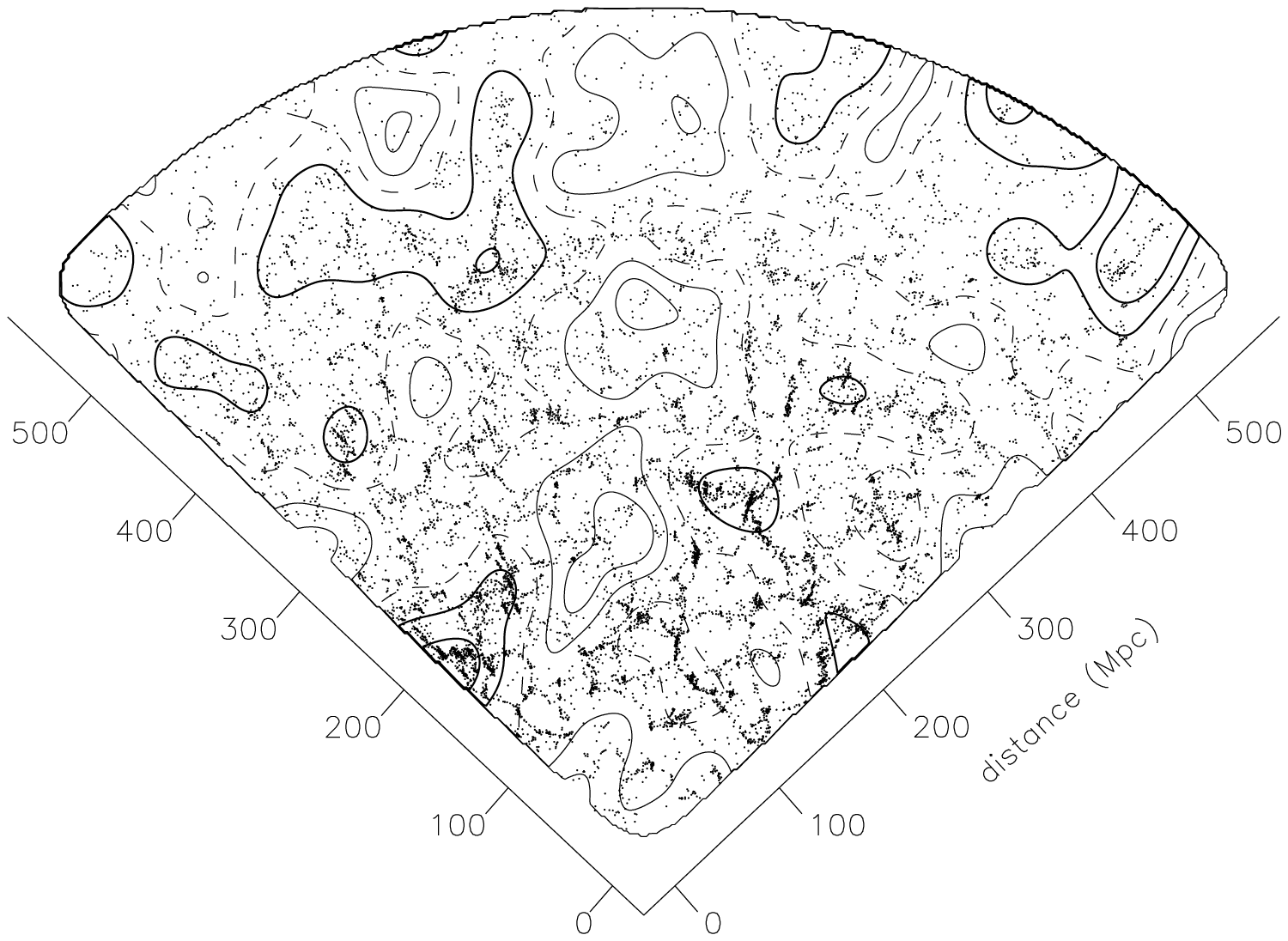}{12cm}{0}{120}{120}{-440}{-430}
\caption{Contours of $\nu = \{-2,-1,0,1,2\}$ in the $\eta = -30\deg$ slice
of the SSDSS, after smoothing with a Gaussian filter $e^{-r^2/2R_s^2}$,
with $R_s = 20\hMpc$.  The median density contour ($\nu = 0$) is dashed.
High density contours are heavy and solid ($\nu = \{1,2\}$); low density
contours are light and solid ($\nu = \{-2,-1\}$).  Over-plotted are the galaxy
locations themselves, so that the location of overdense regions and voids is
obvious.}
\label{SSDSS.slice}
\end{figure}

\begin{figure}[p]
\plotfiddle{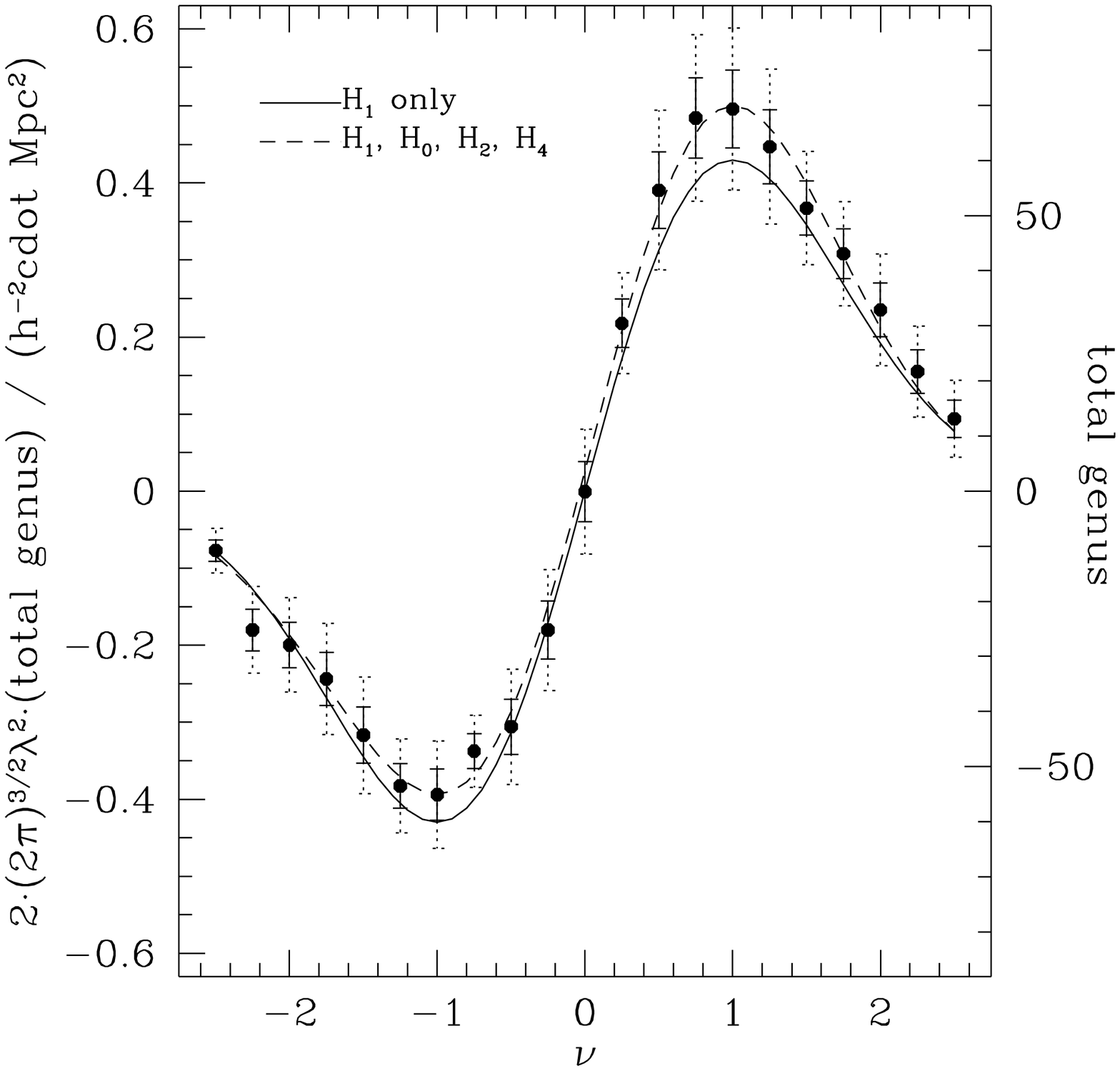}{13cm}{0}{80}{80}{-240}{-110}
\caption{The genus curve of two-dimensional structure in SSDSS slices.
The smoothing length is $R_s=20\hMpc$.  The solid curve is the best fit for
$g = A\cdot\nu\exp(-\nu^2/2)$, expected for a Gaussian random phase field.  The
dashed curve is the best fit for $g = A[H_1(\nu) + BH_0(\nu) + CH_2(\nu) +
DH_4(\nu)]\exp(-\nu^2/2)$, where $H_n$ is the Hermite polynomial of degree
$H_n$ [$H_1(\nu) = \nu$].  Points are the average results from six $1.5\deg$
slices; solid and dotted error bars show the 68.3\% and 95\% confidence
intervals on each $g(\nu)$ measurement computed using the 
Student's $t$-distribution and the variance among the six slices.
The $y$-axis is also labeled (on the right) by the total genus
in the sample.}
\label{SSDSS.2dgen}
\end{figure}

\begin{figure}[p]
\plotfiddle{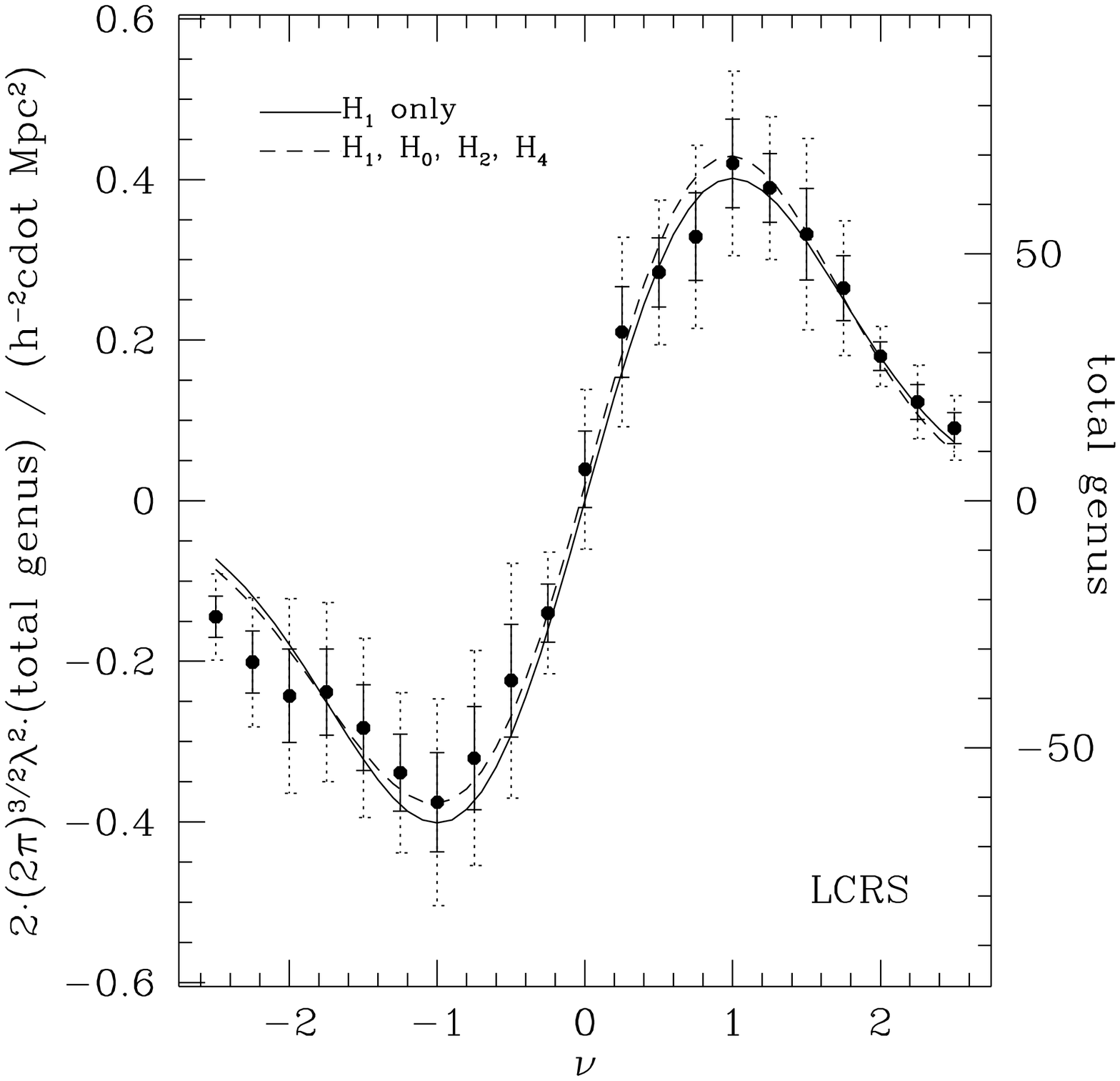}{13cm}{0}{80}{80}{-240}{-110}
\caption{As in Fig.~\ref{SSDSS.2dgen}, but for the Las Campanas Redshift
Survey.  Notice that the fit is not much improved after adding even terms, as
expected when non-linear structure growth is important.  In the Simulated SDSS,
however, the fit improved substantially.}
\label{SSDSS.lcrsgen}
\end{figure}

\begin{references}
 
\reference{} Adler, R.J. 1981, The Geometry of Random Fields, Wiley, New
York

\reference{}Bahcall, N. A., Fan, X., \& Cen, R. 1998, \apj, 485, L53

\reference{}Bahcall, N., \& Soneira, R.M. 1983, \apj, 270, 20

\reference{}Bardeen, J.M., Bond, J.R., Kaiser, N., \& Szalay, A.S. 1986, \apj,
304, 15

\reference{}Bardeen, J.M., Steinhardt, P.J., \& Turner, M.S. 1983,
Phys. Rev. D., 28, 679

\reference{}Bertschinger, E., et al. 1990, \apj, 364, 370

\reference{}Blanchard, A., \& Bartlett, J. G. 1998, A\&A, 332, L49

\reference{}Broadhurst, T.J., Ellis, R.S., Koo, D.S., \& Szalay, A.S. 1990, 
Nature, 343, 726

\reference{}Bucher, M., Goldhaber, A.S. \& Turok, N. 1995a, Phys. Rev. D., 52,
3314

\reference{}Bucher, M., Goldhaber, A. S. \& Turok, N. 1995b, Phys. Rev. D., 52,
5538

\reference{}Canavezes, A., Springel, V., et al. 1998, \mnras, 297, 777

\reference{}Carlberg, R. G., Yee, H. K. C., Ellingson, E., Abraham, R., 
Gravel, P., Morris, S., \& Pritchet, C. J. 1996, \apj, 462, 32

\reference{}Carlberg, R. G., Yee, H. K. C., \& Ellingson, E. 1997,
\apj, 478, 462

\reference{}Cen, R., \& Ostriker, J. P. 1992, \apj, 399, L113

\reference{}Chiba, M., \& Yoshii, Y. 1999, \apj, 510, 42

\reference{}Cole, S., Hatton, S., Weinberg, D. H., \& Frenk, C. S. 1998, 
\mnras, 300, 945

\reference{}Cole, S., Weinberg, D. H., Frenk, C. S., \& Ratra, B. 1997,
\mnras, 289, 37

\reference{}Coleman, G. D., Wu, C.-C., \& Weedman, D. W. 1980, \apjs, 43, 393

\reference{}Coles, P. 1988, \mnras, 234, 509

\reference{}Coles, P., Davies, A. G., \& Pearson, R. C. 1996, \mnras, 281, 1375

\reference{}Coles, P., Moscardini, L., Plionis, M., Lucchin, F., 
Matarrese, S., \& Messina, A. 1993, \mnras, 260, 572

\reference{}Colless, M. 1998, in Wide Field Surveys in Cosmology,
14th IAP Meeting, (Editions Frontieres: Paris), p. 77

\reference{}Colley, W. N. 1997, \apj, 489, 471

\reference{}Colley, W. N., Gott, J. R., \& Park, C. 1996, \mnras, 281, L82

\reference{}de Lapparent, V., Geller, M., \& Huchra, J. 1986, \apj, 302, L1

\reference{}de Vaucouleurs, G. 1948, Ann d'Astrophys, 11, 247

\reference{}Dekel, A., Bertschinger, E., \& Faber, S. M. 1990, \apj, 364, 349

\reference{}Djorgovski, S., \& Davis, M. 1987, \apj, 313, 59

\reference{}Doroshkevich, A. G. 1970, Astrophysika, 6, 320

\reference{}Dressler, A., et al 1987, \apj, 313, 42

\reference{}Eke, V. R., Cole, S., Frenk, C. S., \& Henry, J. P.  1998,
\mnras, 298, 1145

\reference{}Evrard, A. E. 1997, \mnras, 292, 289

\reference{}Faber, S. M., \& Jackson, R. E. 1976, \apj, 204, 668

\reference{}Freeman, K. C. 1970, \apj, 160, 811

\reference{}Fukugita, M., Futamase, T., Kasai, M., \& Turner, E. L. 1992, \apj,
393, 3

\reference{}Fukugita, M., Ichikawa, T., Gunn, J. E., Doi, M., Shimasaku, K., 
\& Schneider, D. P. 1996, \aj, 111, 1748

\reference{}Geller, M. J. \& Huchra, J. P. 1989, Science, 246, 897 

\reference{}Gooding, A. K., Park, C., Spergel, D. N., Turok, N., \& Gott, J.R.,
1992, \apj, 393, 42

\reference{}Gott, J. R. 1982, Nature, 295, 304

\reference{}Gott, J. R. 1986, Inner Space/Outer Space, The Interface Between
Cosmology \& Particle Physics, eds. E. W. Kolb et al. (Chicago: Univ. of
Chicago Press), 362

\reference{}Gott, J. R., \etal\ 1989, \apj, 340, 625

\reference{}Gott, J. R., Gao, B., \& Park, C. 1991, \apj, 383, 90

\reference{}Gott, J. R., Mao, S., Park, C., Lahav, O. 1992, \apj, 385, 26 

\reference{}Gott, J. R., Melott, A. L., \& Dickinson, M. 1986, \apj, 306, 341
(GMD)

\reference{}Gott, J. R., Park, C., Juskiewicz, R., Bies, W. E., Bennett,
D. P., Bouchet, F. R., \& Stebbins, A. 1990, \apj, 352, 1

\reference{}Gott, J. R., Park, M., \& Lee, H. M. 1989, \apj, 338, 1

\reference{}Gott, J. R., Rhoads, J. E., \& Postman, M. 1994, \apj, 421, 1

\reference{}Gott, J. R. \& Statler, T. S. 1984, Physics Letters, 136B, 157

\reference{}Gott, J. R., Weinberg, D. H. \& Melott, A. L. 1987, \apj, 321, 2
(GWM)

\reference{}Gunn, J. E., et al.\ 1998, \aj, 116, 3040

\reference{}Gunn, J. E., \& Knapp, G. R. 1993, in Sky Surveys: Protostars
to Protogalaxies, ASP Conference Series vol. 43,
ed. B. T. Soifer, (ASP: San Francisco), p. 267

\reference{gw95} Gunn, J. E. \& Weinberg, D. H. 1995, in Wide Field
Spectroscopy and the Distant Universe, eds. S. Maddox \& A. Arag\'on-Salamanca,
(Singapore: World Scientific), 3, astro-ph/9412080

\reference{}Guth, A. H. 1981, Phys. Rev. D. 23, 347

\reference{}Hamilton, A. J. S., Gott, J. R., \& Weinberg, D. W. 1986, \apj,
309, 1 

\reference{}Harrison, E. R. 1970, Phys. Rev. D., 1, 2726

\reference{}Hawking, S., \& Turok, N. 1998, hep-th/9802030

\reference{}Hoessel, J. G. 1980, \apj, 241, 493

\reference{}Kaiser, N. 1987, \mnras, 227, 1

\reference{}Katz, N., Hernquist, L., \& Weinberg, D. H. 1992, \apj, 399, L109

\reference{}Kauffmann, G., Nusser, A., \& Steinmetz, M. 1997, \mnras, 286, 795

\reference{}Kent, S. M. 1985, \apjs, 59, 115

\reference{}Klypin, A. A., \& Kopylov, A. I. 1983, Sov Ast Lett, 9, 41

\reference{}Knapp, G. R., et al.\ 1997, proposal to NASA by the SDSS
collaboration, available at http://www.astro.princeton.edu/BBOOK/

\reference{} Kogut, A., Banday, A. J., Bennett, C. L., Gorski, K. M.,
Hinshaw, G., Smoot, G. F., \& Wright, E. L, 1996, \apj, 464, L29

\reference{}Lauer, T. R., \& Postman, M. 1994, \apj, 425, 418

\reference{}Lin, H., Kirshner, R. P., Shectman, S., Landy, S. D., Oemler, A.,
Tucker, D. L., \& Schechter, P. 1996, \apj, 471, 617

\reference{}Linde, A. D. 1990, Inflation and Quantum Cosmology (Academic Press,
Boston)

\reference{}Linde, A. D. 1995, Phys. Lett. B., 351, 99

\reference{}Linde, A., \& Mezhlnmian, A. 1995, Phys. Rev. D., 52, 6789

\reference{}Lineweaver, C. H. 1998, \apj, 505, L69

\reference{}Lupton, R. H. 1993. Statistics in Theory and Practice
Princeton University Press, Princeton, NJ, pages 27ff

\reference{}Lynden-Bell, D., et al. 1988, \apj, 326, 618

\reference{}Maddox, S. J., Efstathiou, G., Sutherland, W. J., \& Loveday,
J. 1990, \mnras, 242, 43P

\reference{}Maoz, D. \& Rix, H. 1993, \apj, 416, 425

\reference{}Margon, B. 1999, Phil Trans Roy Soc Lond A, 357, in press,
astro-ph/9805314

\reference{}Matsubara, T. 1994, \apjl, 434, 43

\reference{}Matsubara, T. 1996, \apj, 457, 13

\reference{}Matsubara, T. \& Suto, Y., 1996, \apj, 460, 51

\reference{}Melott, A. L., Cohen, A. P., Hamilton, A. J. S., Gott, J. R. \&
Weinberg, D. H. 1989, \apj, 345, 618

\reference{}Melott, A. L., \& Dominik, K. G. 1993, \apjs, 86, 1

\reference{}Melott, A. L., Weinberg, D. H., \& Gott, J. R. 1988, \apj, 328, 50

\reference{}Moore, B., Frenk, C. S., Weinberg, D. H., Saunders, W.,
Lawrence, A., Ellis, R. S., Kaiser, N., Efstathiou, G., \& Rowan-Robinson, M.,
1992, \mnras, 256, 477

\reference{}Narayanan, V. K., Berlind, A. A., \& Weinberg, D. H. 1998, \apj,
submitted, astro-ph/9812002

\reference{}Oke, J. B., \& Gunn, J. E. 1983, \apj, 266, 713

\reference{}Ostriker, J. P., \& Steinhardt, P. J. 1995, Nature, 377, 600

\reference{}Park, C. 1990, PhD Thesis, Princeton University

\reference{}Park, C. 1990, \mnras, 242, 59P

\reference{}Park, C. 1991, \mnras, 251 167

\reference{}Park C., Colley, W. N., Gott, J. R., Ratra, B., Spergel, D. N., \&
Sugiyama, N. 1998, \apj, 506, 473

\reference{}Park, C., \& Gott, J. R. 1991a, \mnras, 249, 288

\reference{}Park, C., \& Gott, J. R. 1991b, \apj, 378, 457

\reference{}Park, C., Gott, J. R., \& da Costa, L. N. 1992, \apj, 392, L51

\reference{}Park, C., Gott, J. R., Melott, A., \& Karachentsev, I. D. 1992,
\apj, 387, 1

\reference{}Peacock, J. A., \& Dodds, S. J. 1994, \mnras, 267, 1020

\reference{}Peebles, P. J. E. 1993, Principles of Physical Cosmology,
(Princeton: Princeton University Press)

\reference{}Peebles, P. J. E., \& Ratra, B, \apj 325, L17

\reference{}Peebles, P. J. E., \& Yu, J. T. 1970, \apj, 162, 815

\reference{}Perlmutter, S., et al. 1999, \apj, in press, astro-ph/9812133

\reference{}Petrosian, V. 1976, \apj, 209, L1

\reference{}Plionis, M., Valdarnini, R., \& Coles, P. 1992, \mnras, 258, 114

\reference{}Protogeros, Z. A. M., \& Weinberg, D. H. 1997, \apj, 489, 457

\reference{}Ratra, B., \& Peebles, P. J. E. 1994, \apj, 432, L5

\reference{}Ratra, B., \& Peebles, P. J. E. 1995, Phys. Rev. D52, 1837

\reference{}Ratra, B., \& Peebles, P. J. E. 1998, Phys. Rev. D37, 3406

\reference{}Reichart, D. E., Nichol, R. C., Castander, F. J., Burke, D. J.,
Romer, A. K., Holden, B. P., Collins, C. A., \& Ulmer, M. P. 1998,
\apj, submitted, astro-ph/9802153

\reference{}Riess, A. G., et al.\ 1998, \aj, 116, 1009

\reference{}Roos, M., \& Harun-or-Rashid, S. M. 1999, A\&A, submitted, 
astro-ph/9901234

\reference{}Ryden, B. S. 1992, \apj, 396, 445

\reference{}Saunders, W. et al. 1991, Nature, 349, 32

\reference{}Schechter, P. 1976, \apj, 203, 297

\reference{}Schmidt, M. 1968, \apj, 151, 393

\reference{}Shandarin, S. F., \& Zel'dovich, Y. B. 1989, Rev Mod Phys, 61, 185

\reference{}Shane, C. D., \& Wirtanen, C. A. 1967, Pub Lick Obs, 22, part 1

\reference{}Shectman, S. A., Landy, S. D., Oemler, A., Tucker, D. L., Lin, H.,
Kirshner, R. P., \& Schechter, P. L. 1996, \apj, 470, 172

\reference{}Smoot, G. F. \etal\, 1994, \apj 437, 1

\reference{}Smoot, G. F. et al. 1992, \apj, 396, L1

\reference{}Springel, V., et al.\ 1998, \mnras, 298, 1169

\reference{}Tegmark, M. 1998, \apj, submitted, astro-ph/9809201

\reference{}Tully, R. B., \& Fisher, J. R. 1977, \aap, 54, 661

\reference{}Turner, M. S. 1998, in Critical Dialogues in Cosmology, ed. Turok,
N. (Singapore: World Scientific), 555

\reference{}Viana, P. T. P., \& Liddle, A. R. 1998, \mnras, in press, 
astro-ph/9803244

\reference{}Vogeley, M. S., Park, C., Geller, M. J., Huchra, J. P., Gott,
J. R. 1994, \apj, 420, 525

\reference{}Weinberg, D. H. 1988, PASP, 100, 1373

\reference{}Weinberg, D. H., \& Cole, S., 1992, \mnras, 259, 652

\reference{}Weinberg, D. H., Croft, R. A. C., Hernquist, L., 
Katz, N., \& Pettini, M. 1998, \apj, submitted, astro-ph/9810011

\reference{}Weinberg, D. H., \& Gunn, J. E. 1990a, \apj, 352, L25

\reference{}Weinberg, D. H., \& Gunn, J. E. 1990b, \mnras, 247, 260

\reference{}White, S. D. M., Frenk, R. S., Davis, M., \& Efstathiou, G. 1987,
\apj, 313, 505

\reference{}Wright, E. L. et al. 1992, \apj, 396, L13

\reference{}Yamamoto, K., Sasaki, M., \& Tanaka, T. 1995, \apj, 455,412

\reference{}Zel'dovich, Y. B. 1970, A\&A, 5, 84

\reference{}Zeldovich, Ya. B. 1972, \mnras, 160, 1P

\end{references}
\end{document}